 \documentclass[aps, twocolumn, nopacs, amsmath]{revtex4}
\usepackage{dcolumn}
\usepackage{bm}
\usepackage{graphicx}
\usepackage{color}
\usepackage{subfigure}
\usepackage{hyperref}
\usepackage{latexsym}
\usepackage{amsthm}
\usepackage{amssymb}
\usepackage{array}
\usepackage{braket}
\DeclareGraphicsExtensions{.jpg,.pdf, .mps, .png, .eps, .ps, .EPS,.gif}

\begin{document}
\def\be{\begin{equation}}
\def\ee{\end{equation}}

\def\bc{\begin{center}}
\def\ec{\end{center}}
\def\bea{\begin{eqnarray}}
\def\eea{\end{eqnarray}}
\newcommand{\avg}[1]{\langle{#1}\rangle}
\newcommand{\Avg}[1]{\left\langle{#1}\right\rangle}

\def\ie{\textit{i.e.}}
\def\etal{\textit{et al.}}
\def\m{\vec{m}}
\def\G{\mathcal{G}}

\newcommand{\davide}[1]{{\bf\color{blue}#1}}
\newcommand{\gin}[1]{{\bf\color{green}#1}}

\title{Renormalization group theory of percolation on  pseudo-fractal \\simplicial and cell complexes}

\author{Hanlin Sun}
\affiliation{School of Mathematical Sciences, Queen Mary University of London, London, E1 4NS, United Kingdom}
\author{Robert M. Ziff}
\affiliation{Center for the Study of Complex Systems and Department of Chemical Engineering, University of Michigan, Ann Arbor, Michigan 48109-2800, USA}
\author{Ginestra Bianconi}
\affiliation{The Alan Turing Institute, 96 Euston Rd, London NW1 2DB, United Kingdom\\ 
School of Mathematical Sciences, Queen Mary University of London, London, E1 4NS, United Kingdom}

\begin{abstract}
Simplicial complexes are gaining increasing scientific attention as  they are generalized network structures that can represent the many-body interactions existing in complex systems raging from the brain to high-order social networks. Simplicial complexes are formed by  simplicies, such as nodes, links, triangles and so on. Cell complexes further extend these generalized network structures as they are formed by  regular polytopes such as squares, pentagons etc.  Pseudo-fractal simplicial and cell complexes are a major example of generalized network structures and they  can be obtained by gluing $2$-dimensional  $m$-polygons ($m=2$ triangles, $m=4$ squares, $m=5$ pentagons, etc.)  along their links according to a simple iterative rule. Here we investigate the interplay between the topology of pseudo-fractal simplicial and cell complexes and their dynamics by characterizing the critical properties of  link percolation  defined on these structures. By using the renormalization group we show that the pseudo-fractal simplicial and cell complexes have a  continuous percolation threshold at $p_c=0$. When the pseudo-fractal structure is formed by polygons of the same size $m$, the transition is  characterized by an exponential suppression of the order parameter $P_{\infty}$ that depends on the number of sides $m$ of the polygons forming the pseudo-fractal cell complex, i.e., $P_{\infty}\propto p\exp(-\alpha/p^{m-2})$. Here these results are also generalized to random pseudo-fractal cell-complexes formed by polygons of different number of sides $m$.
\end{abstract}

\maketitle

\section{Introduction}
Simplicial and cell complexes \cite{Perspectives,Lambiotte} are generalized network structures capturing the many-body interactions existing in complex systems such as brain networks \cite{Bassett,Petri,BlueBrain}, social networks \cite{Alain,iacopini2019simplicial,Yamir}, and complex materials \cite{Bassett_granular,Tadic}.
Simplicial and cell complexes are  not only formed by nodes and links like networks, but they are also formed  by  higher dimensional simplexes and polytopes such as triangles, squares, pentagons, etc.
Being formed by geometrical and topological building blocks simplicial complexes are ideal structures to study network geometry \cite{Emergent,Hyperbolic,Polytopes}.
Moreover, simplicial complexes are key to  investigate the role that network geometry and  many-body interactions have on dynamics.
Among the vast variety of dynamical processes that are starting to be investigated on simplicial complexes we mention percolation \cite{Bianconi_Ziff,kryven2019renormalization,branching,hasegawa}, synchronization  \cite{Arenas,millan2020explosive,Boccaletti,Battiston_synchronization,millan2018complex,millan2019synchronization}, epidemic spreading \cite{iacopini2019simplicial,laurent}, Gaussian models \cite{bianconi2020spectral,reitz,Tadic}, and random walks \cite{torres,Battiston_RW}.
The vast majority of hierarchical networks studied in statistical mechanics and network theory literature is formed by the skeleton of simplicial and cell complexes (i.e., the network formed by its nodes and links). Examples range from the diamond network of Migdal and Kadanoff \cite{migdal1976phase,kadanoff1976notes} to the hyperbolic Farey graphs that have been  shown to display a  discontinuous percolation phase transition in Ref.\  \cite{hyperbolic_Ziff}.
These networks are well suited to perform exact real-space renormalization group (RG) calculations. Using RG theory there has been very important progress in characterizing the critical properties of percolation \cite{hyperbolic_Ziff,Patchy,RG,flower_tau,tricritical,
kryven2019renormalization,Bianconi_Ziff,branching,percolation_Apollonian}, spin (Ising and Potts) models \cite{Boettcher_RG,Boettcher_Potts,Boettcher_Ising,Berker_RG}, and Gaussian models \cite{bianconi2020spectral,reitz} in these structures.
In particular in Refs.\ \cite{kryven2019renormalization,branching} the robustness of the result obtained by Boettcher, Singh and Ziff in Ref.\  \cite{hyperbolic_Ziff} has been investigated by considering more general simplicial and cell complexes.
It has been found that two-dimensional simplicial and cell complexes, i.e., simplicial and cell complexes build by gluing two-dimensional polygons along their links, can display a large variety of critical behaviors for the order parameter of link percolation.
Here we extend this line of research and we characterize the link percolation transition to random pseudo-fractal simplicial and cell complexes.  Pseudo-fractal simplicial complexes have been originally proposed as deterministic model for complex networks in Ref.\  \cite{dorogovtsev2002pseudofractal}. Link percolation on these deterministic pseudo-fractal networks has been discussed previously in Ref.\  \cite{doro3}. Here however we provide a more extensive treatment of the problem and are able to show that the critical percolation properties of the   deterministic pseudo-fractal simplicial complex differs from the percolation properties of the deterministic pseudo-fractal cell complex and the random pseudo-fractal simplicial complexes.
Indeed our work shows that for the deterministic pseudo-fractal simplicial complexes formed by $m$-polygons, the phase transition is at $p_c=0$ and the order parameter behaves as
\bea
P_{\infty}\propto p\exp(-\alpha/p^{m-2})
\label{sum}
\eea
where $\alpha$ is a constant.
Therefore the exponential suppression goes like  $1/p$ for $m=3$ as obtained by Ref.\ \cite{doro3}
but goes like $1/p^{m-2}$ for $m>3$. On a side note we mention also  that our derivation also captures the factor $p$ in Eq.\ (\ref{sum}) not discussed in Ref.\  \cite{dorogovtsev2002pseudofractal}.
Finally for random pseudo-fractal simplicial complexes we show that the critical behavior is dictated by the smaller value of $m$ of the polygons of the cell complex.
 
 The paper is structured as follows: in Sec.\  II we describe the main properties of the random pseudo-fractal cell complexes studied in this work; in Sec.\  III we introduce link percolation on pseudo-fractal cell complexes, we derive the iterative equations for the linking probability defining the RG equations, and we derive the expression for the generating functions and the for the order parameter in terms of the linking probability, in Sec.\  IV we discuss the RG flow, in Sec.\ V we derive the critical behavior of the order parameter, finally in Sec.\  VI we provide the conclusions.

 \begin{figure*}
  \centerline{\includegraphics[width=1.9\columnwidth]{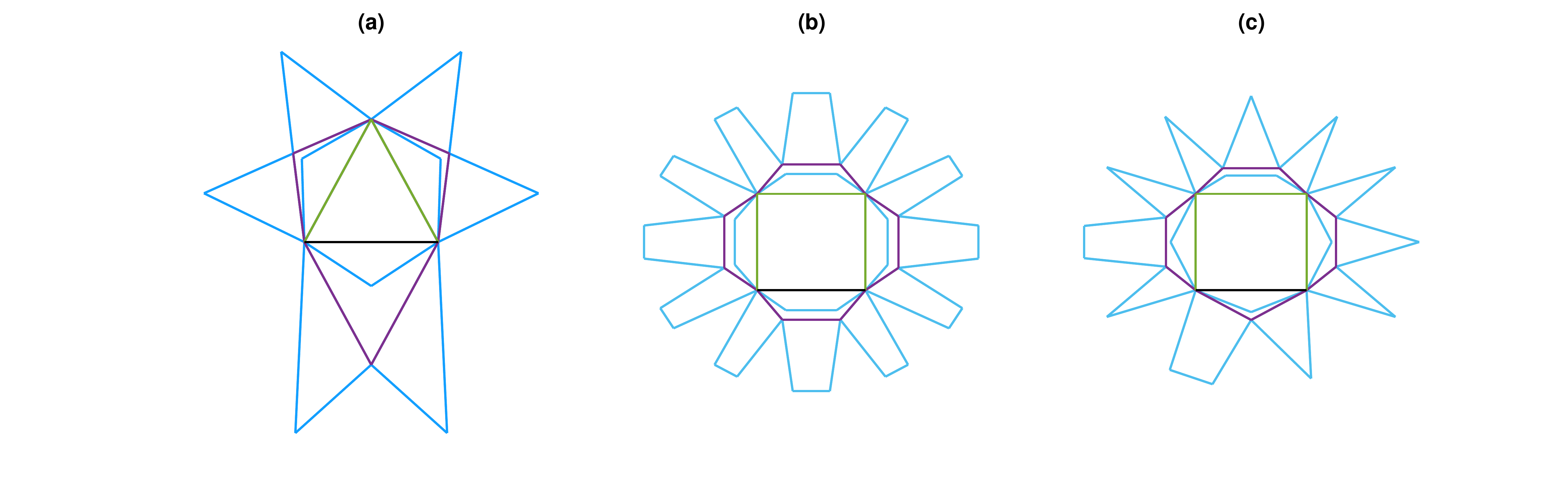}}
  \caption{ (Color online)   Examples of   pseudo-fractal simplicial and cell complexes  represented at iteration $n=3$. Panel (a)  shows a deterministic pseudo-fractal simplicial  complex with $m=3$, panel (b) shows a deterministic pseudo-fractal cell complex with $m=4$, panel (c) shows a random pseudo-fractal cell complex with $q_{3}=q_{4}=1/2$. The different colors indicate the different iterations: $n=0$ (black), $n=1$ (green) $n=2$ (purple), $n=3$ (cyan).}
\label{Fig:cell_complexes}  
\end{figure*}

 \section{Random pseudo-fractal simplicial and cell complexes}
 The pseudo-fractal simplicial complex \cite{dorogovtsev2002pseudofractal} is  constructed iteratively starting at iteration $n=0$ from a single link.
At each time $n\geq 1$ we attach a triangle to every link introduced at iteration $0\leq n^{\prime}<n$.
This construction can be generalized by considering a random  cell complex formed by regular $m$-polygons with different $m\geq 3$.
We start at iteration $n=0$ from an initial link. 
At each iteration $n\geq 1$ we glue a $m$-polygon to every link of the cell complex introduced at iteration $0\leq n^{\prime}<n$ with  $m\geq 3$  drawn from a $q_m$ distribution.
Is is easy to show that at iteration $n$ the  expected number of nodes $\overline{N}_n$ and links $\overline{L}_n$  are given by
\bea
\overline{N}_n&=&2+\frac{\avg{m}-2}{\avg{m}-1}(\avg{m}^{n}-1),\nonumber \\
\overline{L}_n&=&\avg{m}^n,
\eea
where $\avg{m}=\sum_{m\geq 3}mq_m$.
In the following we will refer to these generalized network structures as random cell complexes.
However for $q_m=\delta_{m,3}$ the random pseudo-fractal cell complex reduces to the pseudo-fractal simplicial complex (see Fig.\  \ref{Fig:cell_complexes}a). 
Moreover for $q_{m^{\prime}}=\delta_{m^{\prime},m}$ and $m>3$ we obtain a deterministic cell complex formed by gluing only $m$-polygons.
(see Fig.\  \ref{Fig:cell_complexes}b for an example of a deterministic cell complex with $m=4$). 
Only if the distribution $q_m$ is not a Kronecker delta, the model reduces to a genuine random cell complex (see Fig.\  \ref{Fig:cell_complexes}b for an example of a random  cell complex with $q_3=q_4=1/2$).

\section{Link percolation on pseudo-fractal simplicial and cell complexes }

\subsection{Link probability}
In this paper we investigate the critical properties of link percolation on pseudo-fractal cell complexes. We assume that each link is retained with probability $p$. It follows that each link is removed with probability $q=1-p$.
In order to study link percolation on pseudo-fractal cell complexes we first derive the RG equations for the linking probability $T_n$ that the two initial nodes of the pseudo-fractal cell complex are linked at iteration $n$.
At iteration $n=0$ the two initial nodes are connected if the link between them is present, therefore $T_0=p$.
At iteration $n\geq 0$ the two initial nodes are connected by a path except if the initial link is not present and the two nodes are not connected by any path passing through any of the  $m$-polygons glued to initial link at different iterations.
Therefore for a deterministic pseudo-fractal cell complex with $q_{m^{\prime}}=\delta_{m^{\prime},m}$ we obtain
\bea
T_{n+1}=1-(1-p)\prod_{j=0}^{n}(1-T_j^{m-1})
\eea
with initial condition $T_0=p$. 
For the  random pseudo-fractal cell complexes the iterative equations determining $\{T_n\}_{n\geq 0}$ needs to take into account the randomness of $m$. It is therefore immediate to show that we have
\bea
T_{n+1}=1-(1-p)\left[\prod_{j=0}^n(1-Q(T_j))\right],
\label{pseudo_T}
\eea
where $Q(T)$ is given by 
\bea
Q(T)=\sum_{m\geq 3}q_m T^{m-1},
\label{Q}
\eea
with $T_0=p$.
This recursive set of equations can be also written as 
\bea
T_{n+1}=1-(1-T_n)(1-Q(T_n)),
\label{RGflow}
\eea
with $T_0=p$.
This equation can be also derived directly, without making use of Eq.\ (\ref{pseudo_T}) as it implies that the two initial nodes are connected  at iteration $n+1$ unless they are not connected  at iteration $n$ (that happens with probability $1-T_n$) and they are neither connected by the polygon added at iteration $n=1$ (that happens with probability $1-Q(T_n)$).
The fixed point solutions are only 
\bea
T^{\star}=0,\ \ \ 
T^{\star}=1.
\eea
For any $p>0$ the recursive equations go to the fixed point $T^{\star}=1$.
Instead exactly at $p=0$ the steady state solution is $T^{\star}=0$.
Therefore for any link probability $p>0$ the percolation probability of an infinite network is $T^{\star}=1$. Indeed the RG flow described by Eq.\ (\ref{RGflow}) starts with $T_0=p$ and in the limit $n\to \infty$ reaches 
\bea
\lim_{n\to \infty}T_n=T^{\star}=\left\{\begin{array}{cll}1&\mbox{if}&p>0,\\0&\mbox{if}&p=0.\\
\end{array}.\right.
\eea

 Therefore the (upper) percolation threshold is 
 \bea
 p_c=0.
 \eea
 At $p=p_c=0$ the percolation probability is 
 \bea
 T_c=0.
 \eea
\subsection{Generating function}
In this paragraph we derive the expression for the generating function   $\hat{T}_n(x)$ and $\hat{S}_n(x,y)$ which are key to determine the properties of the link percolation in the pseudo-fractal cell complexes.
The function   $\hat{T}_n(x)$  is the generating function   of the  
number of nodes in the connected component linked to both initial nodes of the considered random branching network. The function  $\hat{S}_n(x,y)$ is  the generating function  of the sizes of the two connected components linked  exclusively to one of the two initial nodes of the same network. 
These generating functions are defined as 
\bea
\hat{T}_n(x)&=&\sum_{\ell=0}^{\infty}t_n(\ell) x^{\ell},\nonumber \\
\hat{S}_n(x,y)&=&\sum_{\ell,\overline{\ell}}s_n(\ell,\overline{\ell})x^{\ell}y^{\overline{\ell}},
\eea
where  $t_n(\ell)$ indicates  the probability that  $\ell$ nodes are  connected to  the two initial nodes and  $s_n(\ell,\overline{\ell})$  indicates  the joint probability that  $\ell$ nodes are connected exclusively to one initial node and  $\overline{\ell}$ nodes are connected exclusively to other initial node.
Therefore for every value of $n$, $t_n(\ell)$ and $s_n(\ell,\overline{\ell})$ obey the normalization condition 
\bea
\sum_{\ell=0}^{\infty}t_n(\ell)+\sum_{\ell=0}^{\infty}\sum_{\overline{\ell}=0}^{\infty}s_n(\ell,\overline{\ell})=1,
\eea
which implies
\bea
\hat{T}_n(1)+\hat{S}_n(1,1)=1.
\eea
\begin{figure}[ht]
  \centerline{\includegraphics[width=0.99\columnwidth]{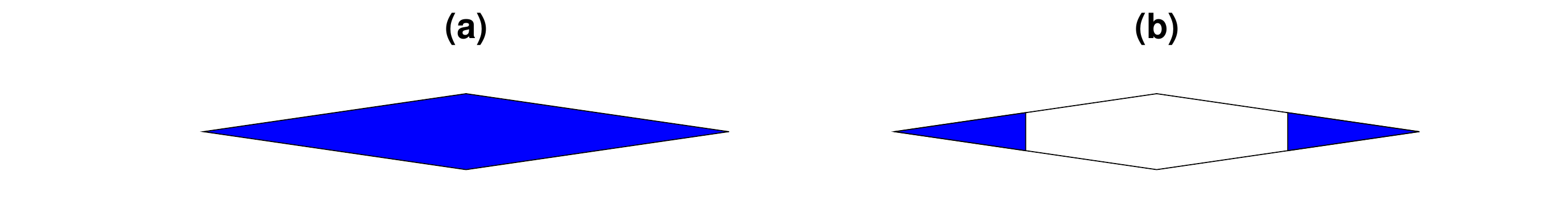}}
  \caption{(Color online) Diagrammatic representation of generating functions $\hat{T}_n(x)$ (a) and $\hat{S}_n(x,y$) (b).   Filled areas indicate connected components that either connect to both end nodes  [$\hat{T}_n(x)$]  or   connect  to a single end node [$\hat{S}_n(x,y)$]. }
\label{Fig:diagram_def}  
\end{figure}
\begin{figure*}[ht]
  \centerline{\includegraphics[width=1.79\columnwidth]{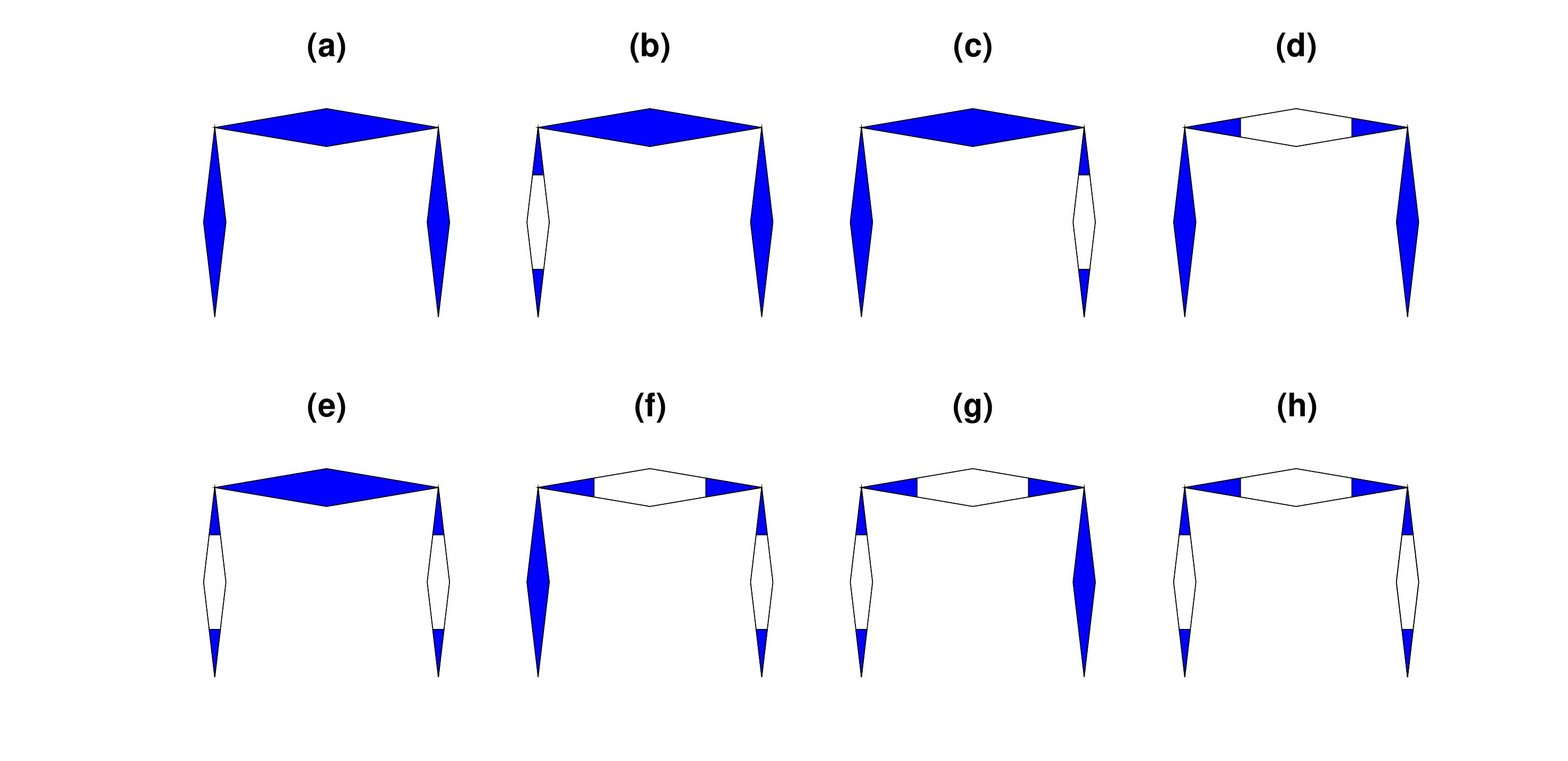}}
  \caption{(Color online) 
  The diagrams coming from a single $m$-polygon added at iteration $n+1-j$  with $m=4$ are shown. These diagrams represent terms that contribute to $\hat{T}_{n+1}(x)$ and $\hat{S}_{n+1}(x,y)$. Diagram (a) contributes exclusively to $\hat{T}_{n+1}(x)$ with a term $x^2\hat{T}^3_j$, diagrams (b)-(h) contribute both to  $\hat{T}_{n+1}(x)$ and $\hat{S}_{n+1}(x,y)$.  The diagrams (b)-(h) contribute to $\hat{T}_{n+1}(x)$ under the assumption that both initial nodes are connected either by the initial link or by polygons added at different generations. The contribution of the diagrams (b)-(h) to $\hat{T}_{n+1}(x)$ are: (b), (c) and (d) $x^2\hat{T}^2_j(x)\hat{S}_n(x,x)$; (e) and (h)  $\hat{S}^2_j(x,1)$; (f) and (g) $x\hat{T}_j(x)\hat{S}^2_j(x,1)$.   The diagrams (b)-(h) contribute to $\hat{S}_{n+1}(x,y)$ under the assumption that the two  initial nodes not  connected  by the initial link and  by any other polygon added at different generations. The contribution of the diagrams (b)-(h) to $\hat{S}_{n+1}(x,y)$ are: (b) $y^2\hat{T}^2_j(y)\hat{S}_n(x,y)$; (c) $x^2\hat{T}^2_j(x)\hat{S}_n(x,y)$;  (d) $xy\hat{T}_j(x)\hat{T}_j(y)\hat{S}_n(x,y)$; (e) and (h)  $\hat{S}_j(x,1)\hat{S}_j(y,1)$; (f),$x\hat{T}_j(x)\hat{S}_j(x,1)\hat{S}_j(y,1)$; (g) $y\hat{T}_j(y)\hat{S}_j(x,1)\hat{S}_j(y,1)$.}
\label{Fig:diagram}  
\end{figure*}
The   generating functions at iteration $n=0$ are given by  
\bea
\hat{T}_{0}(x)&=&p\nonumber \\
\hat{S}_{0}(x,y)&=&1-p,
\eea because initially the two nodes can be either connected by a link (which occurs with probability $p$) or not connected by a link (which occurs with probability $1-p$). In both cases the two initial nodes are not connected to any other node so $t_n(0)=p$,  and $t_n(\ell)=0, $ for all $\ell>0$; similarly $s_n({0,0})=1-p$  and $s_n(\ell, \overline{\ell})=0$ for all $(\ell,\overline{\ell})\neq (0,0)$.

Our aim is to write a set of recursive equations for $\hat{T}_{n+1}(x)$ and $\hat{S}_{n+1}(x,y)$ expressing the generating functions at iteration $n+1$ given the expression of the generating functions at previous generations. To this end we follow the  diagrammatic representation of the generating functions $\hat{T}_n(x) $ and $\hat{S}_n(x,y)$, already introduced   in Refs.\  \cite{hyperbolic_Ziff,kryven2019renormalization,branching}. In particular we represent $\hat{T}_n(x)$ and $\hat{S}_n(x,y)$ with the diagrams presented in Fig.\  $\ref{Fig:diagram_def}$.

 At iteration $n+1$ the initial link will be incident to $n+1$ polygons  added subsequently at each iteration. The polygon added at iteration $n+1-j$ with $0\leq j\leq n$ has links whose statistical properties are equivalent to the one of the initial link at iteration $j$.
If we consider a single polygon added at iteration $n+1-j$, its links will connect the two initial nodes  to other nodes of the cell complex added at later generations, and these nodes will not be reachable by following links that branch out from other polygons. The polygon added at iteration $n+1-j$ will contribute to the generating functions $\hat{T}_{n+1}(x)$ and $\hat{S}_{n+1}(x,y)$ with terms that can be expressed diagrammatically as described in Fig.\  $\ref{Fig:diagram}(a)$ for a $m$-polygon with $m=4$. Only one of these diagrams, i.e., the diagram corresponding to $x^{m-2}T_j^{m-1}$ (diagram (a) in Fig.\  $\ref{Fig:diagram}$) will guarantee connectivity of the two end nodes.  
Therefore the diagram in Fig.\  $\ref{Fig:diagram}(a)$ and its counterpart diagrams for polygons of different number of sides, cannot contribute to $\hat{S}_{n}(x,y)$. However since the initial link at iteration $n$ is connected to $n$ polygons and connectivity can be  guaranteed by the initial link or, when this link is removed, by any one of the  polygons connected to the initial link, all diagrams contribute to $\hat{T}_{n+1}(x)$.

 In order to calculate the generating function $\hat{S}_n(x,y)$ we need to impose that the initial nodes are not directly connected, i.e., for every polygons we need to consider only the contributions from diagrams that do not guarantee connectivity (diagrams (b)-(h) of Fig.\  $\ref{Fig:diagram}$). 
In this way, for a deterministic pseudo-fractal cell complex we obtain 
\begin{widetext}
\bea
&&\hspace{-5mm}\hat{S}_{n+1}(x,y)=(1-p)\prod_{j=0}^n\left[\sum_{r=0}^{m-2}x^ry^{m-2-r}\hat{T}_j^{r}(x)\hat{S}_j(x,y)\hat{T}_j^{m-2-r}(y)+\sum_{s=0}^{m-3}\sum_{r=0}^{s}x^ry^{s-r}\hat{T}_j^{r}(x)\hat{S}_j(x,1)\hat{S}_j(y,1)\hat{T}_j^{s-r}(y)\right].
\eea
\end{widetext}
The derivation of the recursive equation for $\hat{T}_{n+1}(x)$ is slightly more complex. In fact, in order to guarantee that $\hat{T}_{n+1}(x)$ is the generating function of the connected component connected to both initial nodes, we need to impose connectivity.  As noted before, it is sufficient that the initial link guarantees connectedness or, when this link is removed, it is sufficient that a single polygon contributes for the connectedness of the two initial nodes.  Therefore  we express $T_{n+1}(x)$ as the difference between two terms. The first term considers, for each polygon the contribution of all diagrams (the one that guarantee connectedness and the one that do not). The second term considers for each polygons only the terms that do not guarantee connectedness, i.e. removes from the first term the contribution coming from disconnected configurations. In this way for a deterministic pseudo-fractal cell complex we obtain, 
\begin{widetext}
\bea
&&\hspace{-5mm}\hat{T}_{n+1}(x)=\prod_{j=0}^n\left[x^{m-2}\hat{T}_j^{m-1}(x)+(m-1)x^{m-2}\hat{T}_j^{m-2}(x)S_j(x,x)+\sum_{s=0}^{m-3}(s+1)x^s\hat{T}_j^{s}(x)\hat{S}_j(x,1)\hat{S}_n(1,x)\right]\nonumber \\
&&\hspace{9mm}-(1-p)\prod_{j=0}^n \left[(m-1)x^{m-2}\hat{T}_j^{m-2}(x)S_j(x,x)+\sum_{s=0}^{m-3}(s+1)x^s\hat{T}_j^{s}(x)\hat{S}_j(x,1)\hat{S}_j(1,x)\right].\nonumber
\eea
\end{widetext}

For a random pseudo-fractal cell-complex we can generalize these equations obtaining   for $\hat{T}_n(x)$ and $\hat{S}_n(x,y)$ the recursion 
 
\begin{widetext}
\bea
&&\hspace{-5mm}\hat{S}_{n+1}(x,y)=(1-p)\prod_{j=0}^n\left\{\sum_{m\geq 3}q_m\left[\sum_{r=0}^{m-2}x^ry^{m-2-r}\hat{T}_j^{r}(x)\hat{S}_j(x,y)\hat{T}_j^{m-2-r}(y)+\sum_{s=0}^{m-3}\sum_{r=0}^{s}x^ry^{s-r}\hat{T}_j^{r}(x)\hat{S}_j(x,1)\hat{S}_j(y,1)\hat{T}_j^{s-r}(y)\right]\right\},\nonumber \\
&&\hspace{-5mm}\hat{T}_{n+1}(x)=\prod_{j=0}^n\left\{\sum_{m\geq 3}q_m\left[x^{m-2}\hat{T}_j^{m-1}(x)+(m-1)x^{m-2}\hat{T}_j^{m-2}(x)S_j(x,x)+\sum_{s=0}^{m-3}(s+1)x^s\hat{T}_j^{s}(x)\hat{S}_j(x,1)\hat{S}_n(1,x)\right]\right\}\nonumber \\
&&\hspace{9mm}-(1-p)\prod_{j=0}^n \left\{\sum_{m\geq 3}q_m\left[(m-1)x^{m-2}\hat{T}_j^{m-2}(x)S_j(x,x)+\sum_{s=0}^{m-3}(s+1)x^s\hat{T}_j^{s}(x)\hat{S}_j(x,1)\hat{S}_j(1,x)\right]\right\},\nonumber
\eea
\end{widetext}
with initial conditions $\hat{T}_{0}(x)=1-\hat{S}_{0}(x,y) =p$.

We are particularly interested in the generating function $\hat{T}_n(x)$ whose derivative calculated for $x=1$ gives the expected size of the giant component.
The generating function $\hat{T}_{n+1}(x)$ depends on the generating functions $\hat{T}_j(x)$ and the functions $\hat{\Sigma}_j(x)=\hat{S}_j(x,x)$, and $\hat{S}_j(x)=\hat{S}_j(1,x)$ at iterations $0\leq j\leq n$.
From the above equations for $\hat{T}_{n+1}(x)$ and $S_{n+1}(x,y)$ we can deduce directly the  set of recursive equations for $\hat{T}_{n+1}(x)$, $\hat{\Sigma}_{n+1}(x)$, and $\hat{S}_{n+1}(x)$ which read
\begin{widetext}
\bea
\hat{T}_{n+1}(x) &=& \prod_{j=0}^n\left\{\sum_{m\geq 3}q_m\left[x^{m-2}\hat{T}_j^{m-1}(x)+ (m-1)  x^{m-2}   \hat{T}^{m-2}_j(x)\Sigma_j(x)+ \left(\sum_{i=0}^{m-3}(i+1) x^i \hat{T}_j^i(x) \right)S_j^2(x)\right]\right\}\nonumber \\
&&-(1-p) \prod_{j=0}^n\left\{\sum_{m\geq 3}q_m\left[(m-1)x^{m-2} \hat{T}_j^{m-2}(x) \Sigma_j(x) +\left(\sum_{i=0}^{m-3}(i+1)x^i \hat{T}_j^i(x)\right) S^2_j(x) \right]\right\},\nonumber \\
\hat{\Sigma}_{n+1}(x) &=&(1-p) \prod_{j=0}^n\left\{\sum_{m\geq 3}q_m\left[(m-1)x^{m-2} \hat{T}_j^{m-2}(x) \Sigma_j(x) +\left(\sum_{i=0}^{m-3}(i+1)x^i \hat{T}_j^i(x)\right) S^2_j(x) \right]\right\},\nonumber \\
\hat{S}_{n+1}(x) &=& (1-p)\prod_{j=0}^n\left\{\sum_{m\geq 3}q_m\left[\left(\sum_{i=0}^{m-2}x^i\hat{T}_j^i(x)\right)S_j(x)\right]\right\}.
\label{g23:eq}
\eea
\end{widetext}
These equations differ significantly from the corresponding equations valid for two-dimensional hyperbolic manifolds \cite{hyperbolic_Ziff,kryven2019renormalization} and for branched simplicial complexes \cite{branching}. In fact these equations for $\hat{T}_{n+1}(x), \hat{\Sigma}_{n+1}(x)$ and $\hat{S}_{n+1}(x)$ depend on the entire RG flow of the process, i.e., their left hand side if a function of all $\hat{T}_j(x)$, $\hat{\Sigma}_j(x)$ and $\hat{S}_j(x)$  all previous  iterations  $j$ with  $0\leq j\leq n$. 

This apparent complication of the obtained equations can be removed by introducing an auxiliary  function $K_{n+1}(x)$ (see for instance a similar trick used for the Gaussian model in Refs.\ \cite{bianconi2020spectral,reitz}). In order to show this let us   rewrite the Eqs.\ (\ref{g23:eq})  as 
\begin{widetext}
\bea
\hat{T}_{n+1}(x) &=&\hat{K}_{n+1}(x)-\hat{\Sigma}_{n+1}(x)\nonumber \\ 
\hat{K}_{n+1}(x)&=&\prod_{j=0}^n\left\{\sum_{m\geq 3}q_m\left[x^{m-2}\hat{T}_j^{m-1}(x)+ (m-1)  x^{m-2}   \hat{T}^{m-2}_j(x)\Sigma_j(x)+ \left(\sum_{i=0}^{m-3}(i+1) x^i \hat{T}_j^i(x) \right)S_j^2(x)\right]\right\}\nonumber \\
\Sigma_{n+1}(x) &=&(1-p) \prod_{j=0}^n\left\{\sum_{m\geq 3}q_m\left[(m-1)x^{m-2} \hat{T}_j^{m-2}(x) \Sigma_j(x) +\left(\sum_{i=0}^{m-3}(i+1)x^i \hat{T}_j^i(x)\right) S^2_j(x) \right]\right\},\nonumber \\
S_{n+1}(x) &=& (1-p)\prod_{j=0}^n\left\{\sum_{m\geq 3}q_m\left[\left(\sum_{i=0}^{m-2}x^i\hat{T}_j^i(x)\right)S_j(x)\right]\right\},
\eea
\end{widetext}
with initial conditions 
$\hat{T}_0(x)=p,\hat{\Sigma}_0(x)=\hat{S}_0(x)=1-p,\hat{K}_0(x)=1$.
This latter system of equations  can be expressed by a set of iterative   equations between the variables at iteration $n$ and the variable at iteration $n+1$, i.e.,
\begin{widetext}
\bea
\hat{T}_{n+1}(x) &=&\hat{K}_{n+1}(x)-\hat{\Sigma}_{n+1}(x)\nonumber \\ 
\hat{K}_{n+1}(x)&=&\hat{K}_{n}(x)\left\{\sum_{m\geq 3}q_m\left[x^{m-2}\hat{T}_n^{m-1}(x)+ (m-1)  x^{m-2}   \hat{T}^{m-2}_n(x)\Sigma_n(x)+ \left(\sum_{i=0}^{m-3}(i+1) x^i \hat{T}_n^i(x) \right)S_n^2(x)\right]\right\}\nonumber \\
\Sigma_{n+1}(x) &=&\Sigma_{n}(x)\left\{\sum_{m\geq 3}q_m \left[(m-1)x^{m-2} \hat{T}_n^{m-2}(x) \Sigma_n(x) +\left(\sum_{i=0}^{m-3}(i+1)x^i \hat{T}_n^i(x)\right) S^2_n(x) \right]\right\},\nonumber \\
S_{n+1}(x) &=& S_n(x) \left\{\sum_{m\geq 3}q_m\left[\left(\sum_{i=0}^{m-2}x^i\hat{T}_n^i(x)\right)S_n(x)\right]\right\}.
\label{gen_final}
\eea
\end{widetext}
As we will see in the next section, this  recursive set of equations will turn out to be particular useful for evaluating the expected size of the giant component.

\subsection{Order parameter}
\label{3c}
The order parameter of link percolation is the fraction of nodes $P_{\infty}$ that in thermodynamic limit belongs to the giant component, i.e.,
 \bea
P_{\infty}&=&\lim_{n\to \infty}\frac{M_n}{\overline{N}_n}.
\eea
where $M_n$ is the expected size of the giant component connected to the two initial nodes of the cell complex.
The value of $M_n$ can be derived from the generating function $\hat{T}_n(x)$ by derivation, i.\,e.,
\bea
M_n=\left.\frac{d\hat{T}_n(x)}{dx}\right|_{x=1}.
\label{Mn}
\eea
In order to obtain $M_n$ we rewrite Eqs.\ (\ref{gen_final}) in terms of the vector  
\bea
{\bf V}_n(x)&=& \left(V_n^1(x),V_n^2(x),V_n^3(x),V^4_n(x)\right)^\top\nonumber \\ &=&\left(\hat{T}_n(x),\hat{K}_n(x),\hat{\Sigma}_n(x),\hat{S}_n(x)\right)^\top,\eea 
as 
\bea\label{eq:poly:Vps}
 {\mathbf V}_{n+1}^s(x) ={\mathbf F_{n}}(\{\mathbf V_{n}(x)\}, x ). 
\eea
By using this notation, we note that the derivative of ${\bf V}_{n+1}(x)$ calculated at $x=1$ follows
\bea
\hspace{-6mm}\left.\frac{d{\mathbf V}_{n+1}(x)}{dx}\right|_{x=1}={\bf J}_{n}\left.\frac{d{ \bf V}_n^{p}(x)}{dx}\right|_{x=1}+\left.\frac{\partial{\bf F}_{n}^s}{\partial x}\right|_{x=1},
\label{Vprimeps}
\eea
where ${\bf J}_{n}$ indicates the Jacobian matrix of the system of 
Eqs.\ (\ref{eq:poly:Vps}). The initial condition of Eq.\ (\ref{Vprimeps})  is  ${\dot{\bf V}}_0={\bf 0}$ obtained by taking into consideration that  the initial nodes are not counted.
In order to evaluate Eqs.\ (\ref{eq:poly:Vps}) we need to provide an explicit expression of the  Jacobian matrix ${\bf J}_{n}$ whose elements are given by 
\bea
[\left.{\bf J}_{n}]_{ij}=\frac{\partial{F}_{n+1}^i}{\partial{V}_n^{j}(x)}\right|_{x=1}.
\eea

Let us we indicate with $T_n=\hat{T}_n(1)$, $\Sigma_n=\Sigma_{n}(1)$ $S_n=\hat{S}_n(1)$ and $K_n=\hat{K}_n(1)$   that by definition satisfy  $T_n=1-S_n=1-\Sigma_n$ and ${K}_n=1$.

By direct calculation of the  Jacobian ${\bf J}_{n}$ we notice that ${\bf J}_{n}$ can be expressed  as a function of $Q(T_n)$ and $H(T_n)$ with $Q(T)$ given by Eq.\  (\ref{Q}) and $H(T)$ given by 
\bea
H(T)&=&\sum_{m\geq 3}q_m\sum_{i=0}^{m-2}T^{i}.
\eea
In fact, by using the following two relations
\bea
&&\left(1-T_{n}\right) \sum_{i=0}^{m-3} i(i+1) T_{n}^{i-1}=\nonumber \\
&&2 \sum_{i=0}^{m-3}(i+1) T_{n}^{i}-(m-1)(m-2) T_{n}^{m-3},
\eea
and
\bea
&&\left(1-T_{n}\right) \sum_{i=0}^{m-3}(i+1) T_{n}^{i}=\sum_{i=0}^{m-2} T_{n}^{i}-(m-1) T_{n}^{m-2},\nonumber \
\eea
{and using $T_n=1-S_n=1-\Sigma_n$, $K_n=1$,}
a direct calculation  show that  ${\bf J}_{n}$ is given by
\begin{widetext}
\bea
{\bf J}_{n}=\left(\begin{array}{cccc} Q^{\prime}(T_n)+2T_n[H(T_n)-Q^{\prime}(T_n)]&1&T_nQ^{\prime}(T_n)-S_nH(T_n)&2T_n[H(T_n)-Q^{\prime}(T_n)]\\
Q^{\prime}(T_n)+2[H(T_n)-Q^{\prime}(T_n)]&1&Q^{\prime}(T_n)&2[H(T_n)-Q^{\prime}(T_n)]\\
2S_n[H(T_n)-Q^{\prime}(T_n)]&0&S_n[H(T_n)+Q^{\prime}(T_n)]&2S_n[H(T_n)-Q^{\prime}(T_n)]\\ S_n[H(T_n)-Q^{\prime}(T_n)]&0&0&2S_nH(T_n)\end{array}\right).
\label{J}
\eea
\end{widetext}
Similarly  the inhomogeneous term can be expressed as
\bea
\frac{\partial {\bf F}_{n}}{\partial x} = \left(\begin{array}{c}(m-2)Q(T_n)+2T_n^2(H(T_n)-Q'(T_n))\\
(m-2)Q(T_n)+2T_n[H(T_n)-Q'(T_n)]\\
2T_nS_n[H(T_n)-Q'(T_n)]\\
T_nS_n[H(T_n)-Q'(T_n)]\end{array}\right).\nonumber
\eea
Since we have now an explicit expression for both ${\bf J}_{n}$ and ${\partial{\bf F}_n}/{\partial x}$,  we can numerically integrate Eq.\ (\ref{eq:poly:Vps}) finding the number of nodes $M_n$ in the giant component of pseudo-fractal cell complexes for any value of $n$ (numerical precision permitting).
However we also want to have some analytical predictions of the critical properties of link percolation.
To this end we  notice that for $n>0$ and $T_n<1$ the non-homogeneous term ${\partial{\bf F}_n}/{\partial x}$  is  subleading with respect to the homogeneous one in Eq.\  (\ref{eq:poly:Vps}). However for $n=0$ the homogeneous term vanishes due to the initial condition $\dot{\bf V}_0={\bf 0}$ so therefore the non-homogeneous term cannot be neglected.
 Therefore   we can express $\dot{\bf V}_{n+1}$ as
\bea
 \dot{\bf V}_{n+1} \simeq {\mathcal A}_n \prod_{n'=1}^n\lambda_{n'}{ \bf u}_n,
 \label{Vp}
\eea
where $\lambda_n$ and ${\bf u}_n$ are the largest eigenvalue and  the  corresponding left eigenvector of the Jacobian matrix  ${\mathbf J}_{n}$ 
and ${\mathcal A}_n$
is given by 
\bea
{\mathcal A}_n= \left(\prod_{n'=2}^n \braket{{\bf v}_{n'}|{\bf u}_{n'-1}} \right)\braket{{\bf v}_1|\dot{\bf V}_1},
\label {A}
\eea
with  $\dot{\bf V}_0={\partial{\bf F}_0}/{\partial x}$ and ${\bf v}_n$, indicating the right eigenvector corresponding to the largest eigenvalue of the Jacobian ${\bf J}_n$.

Using Eq.\  (\ref{J})  we can directly calculate the largest eigenvalue $\lambda_n$ of the Jacobian matrix ${\bf J}_n$ which is given by
\bea
\lambda_n = \frac{1}{2}\left[\hat{K}(T_n) + \sqrt{\hat{\Delta}(T_n)}\right],
\label{lambdan}
\eea
where $\hat{\Delta}(T_n)$ and $\hat{K}(T_n)$ are given by
\bea
\hat{K}(T_n)&=&(1-2T)Q'(T_n) + 2H(T_n)+1,\\ \nonumber
\hat{\Delta}(T_n) &=& \left[\hat{K}(T_n)\right]^2+8(T-1)\left[H^2(T_n)+Q'(T_n)\right].
\eea
Note that for $T_n\to 1$ then $\lambda_n\to \avg{m}$.

The right eigenvector $\bf v_n$ corresponding to the largest eigenvalue of ${\bf J}_{n}$ is given by:
\begin{widetext}
\bea
{\bf v}_n = \frac{1}{\mathcal{C}^R}\left(\begin{array}{c}                      
{\hat{K(T_n)}-4H(T_n)(1-T_n)+\sqrt{\hat{\Delta}(T_n)}}{2(H(T_n)-Q'(T_n))(T_n-1)}\\
{\hat{K(T_n)}-4Q'(T_n)(1-T_n)+\sqrt{\hat{\Delta}(T_n)}}\\
-{4(H(T_n)-Q'(T_n))(T_n-1)}\\
-{2(H(T_n)-Q'(T_n))(T_n-1)}\end{array}\right)
\eea
\end{widetext}
and the corresponding left eigenvector ${\bf u}_n$ is given by:
\begin{widetext}
	\bea
{\bf u}_n^L = \frac{1}{\mathcal{C}^L}\left(\begin{array}{c}                      
{2H^3(T_n)+4H(T_n)Q'(T_n)-2Q'^2(T_n)+H^2(T_n)\left[-1-3Q'(T_n)+2T_nQ'(T_n)+\sqrt{\hat{\Delta}(T_n)}\right]}\\
{2H^2(T_n)-2H(T_n)Q'(T_n)+Q'(T_n)\left[1-Q'(T_n)+2T_nQ'(T_n)+\sqrt{\hat{\Delta}(T_n)}\right]}\\
{-2H^3(T_n)+2Q'^2(T_n)+H^2(T_n)\left[-1+Q'(T_n)+2T_nQ'(T_n)+\sqrt{\hat{\Delta}(T_n)}\right]}\\
4(\left[H(T_n)-Q'(T_n)\right]\left[H^2(T_n)+Q'(T_n)\right]\end{array}\right),
\eea
\end{widetext}
where $\mathcal{C}^R$ and $\mathcal{C}^L$ are normalization constants which guarantee that the right and left eigenvectors have absolute value one. Note that the right and left eigenvectors of ${\bf v}_n$ and ${\bf u}_n$ satisfy by definition
\bea
\langle {\bf v}_n | {\bf u}_n\rangle=1.
\label{lr}
\eea
From Eqs.\  (\ref{Vp}) and  (\ref{Mn}) it follows that the expected number of nodes $M_{n+1}$ in the giant component  can be expressed as
\bea
 {M}_{n+1} \simeq {\mathcal A}_n \prod_{n'=1}^n\lambda_{n'}{ u}_n^1,
\label{Mp}
\eea
where $u_n^1$ indicates the first element of the vector ${\bf u}_n$.

In Sec.\  V we will use Eq.\  (\ref{Mp}) to derive the critical properties of link percolation on the pseudo-fractal cell complexes.

\section{RG flow}

In this section we study the RG flow described by Eq.\ (\ref{RGflow}) that we rewrite here for convenience
\bea
T_{n+1}=1-(1-T_n)(1-Q(T_n)),
\label{RGflow2}
\eea
with initial condition $T_0=p$.
By defining the auxiliary variable
\bea
y_{n}=-\ln (1-T_n)
\eea
the RG flow described by Eq.\  (\ref{RGflow2}) can be written as 
\bea
y_{n+1}=G(y_n)=y_n-\ln (1-Q(1-e^{-y_n})).
\label{RGflow_y}
\eea

For $p\ll1 $, i.e., close to  $p_c=0$ we  can develop Eq.\ (\ref{RGflow_y}) close to $T=T_c=0$, $y_c=0$. Stopping at the first relevant term in the expansion of $y_{n+1}-y_{n}$ we obtain
 \bea
 y_{n+1}-y_{n}=q_{\overline{m}} y_n^{\overline{m}-1},
 \label{ys}
 \eea
 with initial condition $y_0=-\ln (1-p)$.  Note that in Eq.\  (\ref{ys}),  $\overline{m}$ indicates the minimum value of $m$ for which $q_m>0$.
 By going in the continuous limit and substituting $y_n$ with a function $y(n)$,  Eq.\ (\ref{ys})  can be written as
 \bea
 \frac{dy}{dn}=q_{\overline{m}} y^{\overline{m}-1}.
 \eea
 By integrating  this equation from $0$ up to $n$ we get
 \bea
 y=y_0\left[1-n/n_c\right]^{-1/(\overline{m}-2)},
 \label{yone}
 \eea
 with 
  \bea
 n_c=\left[(\overline{m}-2)|\ln (1-p)|^{\overline{m}-2}q_{\overline{m}}\right]^{-1}.
 \label{nc}
 \eea 
 In particular $y$ diverges at a finite value of $n=n_c$. 

From Eq.\  (\ref{yone})  using 
 \bea
 1-T_n\simeq e^{-y_n}
 \eea
we get the  the asymptotic scaling valid for $y\ll1$ and $p\ll1$
 \bea
 1-T_n=(1-p)^{\theta_n}
 \eea
 with 
 \bea
 \theta_n&=&\left[1-(\overline{m}-2)|\ln(1-p)|^{\overline{m}-2}q_{\overline{m}}n\right]^{-1/(\overline{m}-2)}.
 \eea
 For  $n\ll n_c$ we can made a further approximation and express $\theta_n$ as 
 \bea
 \theta_n\simeq\exp\left[q_{\overline{m}}p^{\overline{m}-2}n\right].
 \eea
 Therefore for $n\ll n_c$
 \bea
 y_n=y_0 \exp \left[q_{\overline{m}}p^{\overline{m}-2}n\right],
 \label{yone2}
 \eea
 with $y_0=|\ln (1-p)|$.
 
\begin{figure}[ht]
  \centerline{\includegraphics[width=0.99\columnwidth]{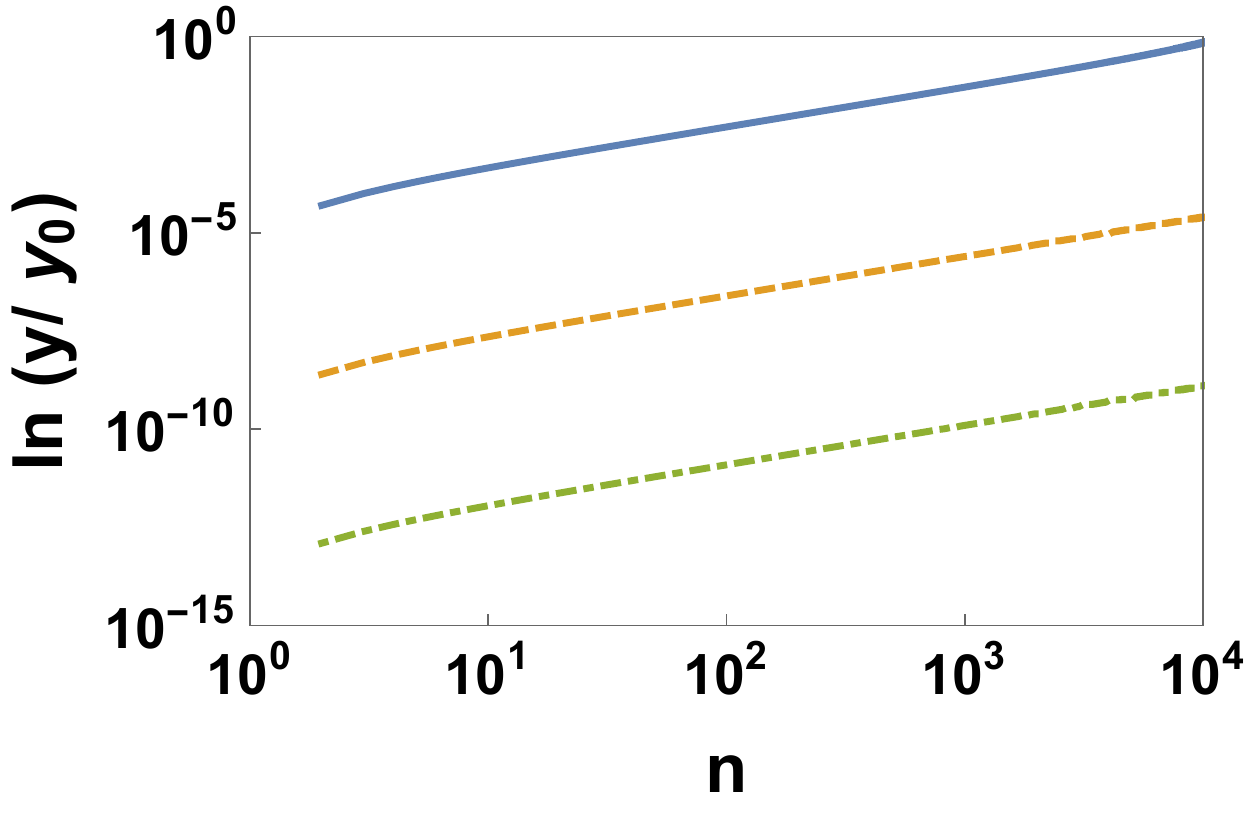}}
  \caption{(Color online) The RG flow is represented by plotting  $\ln(y/y_0)$ (where $y_0=-\ln (1-p)$) versus $n$ for $p=5\times 10^{-5}$ with $m=3$ (blue solid line) $m=4$ (orange dashed line) and $m=5$ (green dot-dashed line). }
\label{Fig:RG_flow}  
\end{figure}

\begin{figure}[ht]
  \centerline{\includegraphics[width=0.99\columnwidth]{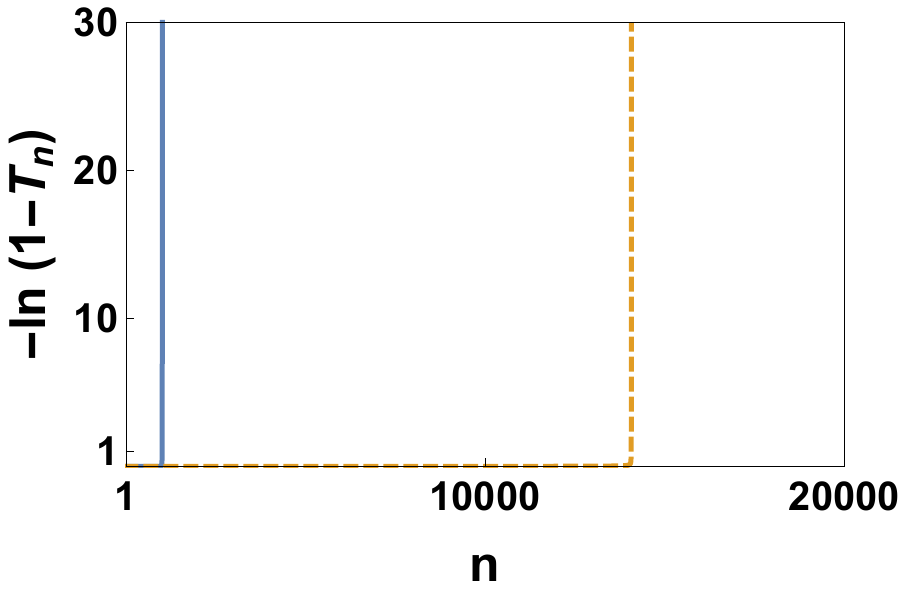}}
  \caption{(Color online) The RG flow is shown by plotting the value $y_n=-\ln (1-T_n)$ where $T_n$ is  the percolation probability, versus $n$ for  fixed value of $p$. The solid (blue) line indicates the RG flow for the deterministic pseudo-fractal simplicial complex with $m=3$ and $p=10^{-3}$,  the dashed (orange) line indicates the RG flow for the deterministic pseudo-fractal cell complex with $m=4$ and $p=6\times 10^{-3}$. The divergence of $y_n$ occurring at a value of $n$ of the order of magnitude of $n_c$ is clearly noticeable, indicating a  discontinuity of $T_n$ reaching the value $T_n=1$ discontinuously. }
\label{Fig:RG_jump}  
\end{figure}
 In Fig.\  \ref{Fig:RG_flow} we show the very good agreement between the numerically integrated value of $y_n$ and the expression given by Eq.\  (\ref{yone2}) for $n\ll n_c$. 
 
Finally  we notice that although Eq.\  (\ref{yone}) is obtained in the limit $y\ll 1$ we can see from numerical integration of the RG flow that $y$  retains  the structure 
 \bea
 y=y_0 f(n/n_c).
 \label{scaling}
 \eea
 Although  the functional form of $f(n/n_c)$ obtained in the expansion for $0<y\ll1$  (which can be deducted from Eq.\  (\ref{yone})) is not exact close to $n\simeq n_c$,  from this expansion we can deduce that $y$ diverges for a finite value of $n$ of the order of $n_c$. In correspondence of this divergence the linking probability $T_n$ jumps to $T_n=1$ (see Fig.\  \ref{Fig:RG_jump}).

 \section{Critical properties of the order parameter}
 
 \subsection{Critical region}
 We are interested in characterizing the properties of the order parameter 
 \bea
P_{\infty}&=&\lim_{n\to \infty}\frac{M_n}{\overline{N}_n}.
\label{Pinf2}
\eea
in the critical region, i.e., close to the percolation threshold  $p_c=0$ taking $0<p\ll1$.
To this end we first discuss the properties of  the expected number of nodes $M_n$ in the giant component when the pseudo-fractal cell complex has evolved up to iteration $n$.
According to the derivation obtained in Sec.\  \ref{3c}, using Eq.\  (\ref{Mp}),  $M_n$ can be approximated   as 
\bea
M_n \simeq {\mathcal A}_{n-1} \prod_{n'=1}^{n-1}\lambda_{n'}{ u}_n^1,
\label{Mp2}
\eea
where ${\mathcal A}_n$ is given by Eq.\  (\ref{A}), which can be written also as
\bea
{\mathcal A}_n={\mathcal D}_n \braket{{\bf v}_1|\dot{\bf V}_1},
\label{A2}
\eea
where ${\mathcal D}_n$ is given by 
\bea
{\mathcal D}_n=\prod_{n'=2}^n \braket{{\bf v}_{n'}|{\bf u}_{n'-1}} 
\eea
For $p\simeq p_c$, ${\mathcal D}_n$ is in first approximation independent of $n$ and approximately equal to one, as the right and left eigenvectors will change slowly with $n$ and by definition  Eq.\  (\ref{lr}) is satisfied.
Therefore Eq.\  (\ref{Mp2}) can be written as 
\bea
M_n \simeq  \braket{{\bf v}_1|\dot{\bf V}_1} \prod_{n'=1}^{n-1}\lambda_{n'}{ u}_n^1.
\label{Mp3}
\eea
\subsection{Critical expansions}

 Our major goal is  to study the critical behavior of the order parameter $P_{\infty}$ (given by Eq.\  (\ref{Pinf2})) depending on the scaling of the expected number of of nodes $M_n$ (whose leading behavior is given by Eq.\ (\ref{Mp3})) in the pseudo-fractal simplicial complex with the number of iterations $n$.  
 
 To this end in this paragraph we will investigate the scaling of  $\lambda_n$ with $n$ for $0<p\ll1$ and we will investigate the scaling of the other factors $\braket{{\bf v}_1|\dot{\bf V}_1} $ and ${ u}_n^1$ present in Eq.\  (\ref{Mp3}) with $p$.   

The leading eigenvalues $\lambda_n$ of the Jacobian matrix ${\bf J}_n$ is expressed according to Eq.\  (\ref{lambdan}) as a function of $H(T_n)$ and $Q(T_n)$.
For $y_n=-\ln (1-T_n)\ll1$ we can expand both $H(T_n)=H(1-e^{-y_n})$ and $Q^{\prime}(T_n)=Q^{\prime}(1-e^{-y_n})$ getting
\bea
Q'(1-e^{-y_n})&=&q_{\overline{m}}(\overline{m}-1)y_n^{\overline{m}-2}+\mathcal{O}(y^{\overline{m}-1}_n,)\nonumber \\
H(1-e^{-y_n})&=&1+y_n+O(y_n^2).
\eea
where $\overline{m}$ indicates the smaller value of $m$ for which $q_m>0$.
Using this expansion in the Eq.\  (\ref{lambdan}) for the maximum eigenvalue $\lambda_n$ of the Jacobian matrix, we get
\bea
\lambda_n=2(1+y_n)+O(y_n^2)
\label{l2}
\eea
For $y_n\ll1$ also the inhomogeneous  term $\partial {\bf F}_{n}/{\partial x}$ can be expanded to give
\bea
\frac{\partial {\bf F}_{n}}{\partial x}\simeq \left(\begin{array}{c} (2+q_{3})y_n^2\\2y_n\\2y_n\\y_n\end{array}\right),
\eea
For $n=0$  where the homogeneous term vanishes due to the trivial initial condition $\dot{\bf V}_0={\bf 0}$ and  the inhomogeneous term has the leading behavior
\bea
\dot{\bf V}_1=\frac{\partial {\bf F}_{0}}{\partial x}\simeq \left(\begin{array}{c} (2+q_{3})p^2\\2p\\2p\\p\end{array}\right).
\eea
Moreover the leading term of ${\bf v}_1$ is 
\bea
{\bf v}_1\simeq\frac{1}{6-4p/3} ((1+p),(1-p),(-1+p),2).
\eea
Therefore for $p\ll 1$ we have that $\braket{{\bf v_1}|\dot{\bf V}_1}$ scales linearly with $p$. In particular
\bea
\braket{{\bf v_1}|\dot{\bf V}_1}\simeq \frac{p}{3}.
\label{braket}
\eea
Finally we observe that for  $n\gg1$ we have $T_n\simeq 1$ and the right eigenvector corresponding to the largest eigenvalue scales like
\bea
{\bf u}_n=(1,1,0,0)^{\top}.
\label{unlim}
\eea
By considering the scaling relations determined by Eq.\  (\ref{braket}) and Eq.\  (\ref{unlim}) in Eq.\  (\ref{Mp3}) we obtain that for $n\gg1$ the fraction of nodes $M_n$ in the giant component obeys
\bea
M_n\propto p \prod_{n'=1}^{n-1} \lambda_{n^{\prime}}.
\label{Mp4}
\eea
with $\lambda_n$ following Eq.\ (\ref{l2}) for $y_n\ll1$.

\begin{figure}
  \centerline{\includegraphics[width=0.90\columnwidth]{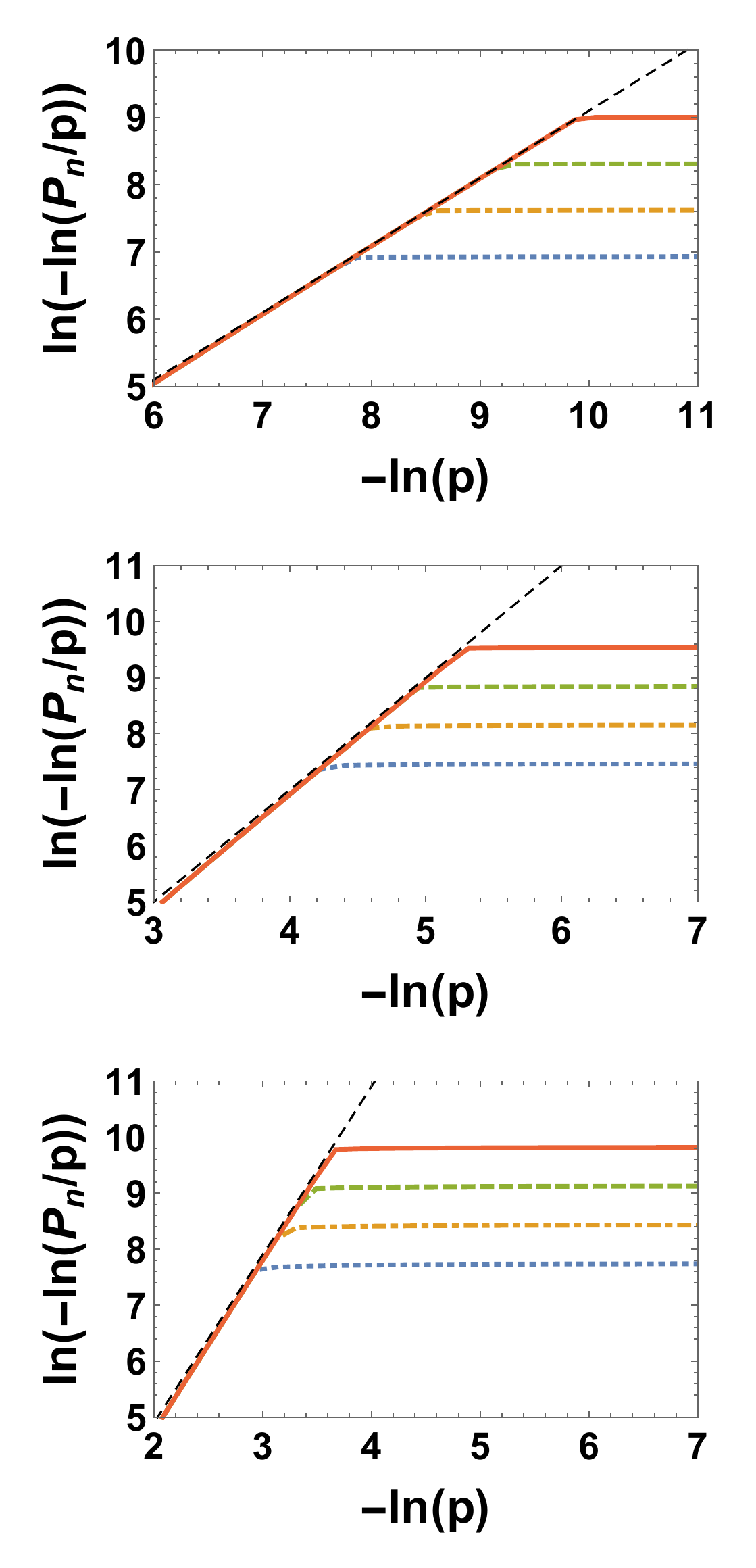}}
  \caption{(Color online) The scaling of the order parameter $P_n$ is shown a as function of $p$ for the deterministic pseudo-fractal simplicial  and cell complexes with  $m=3$ (top panel), $m=4$ (central panel) and $m=5$ (bottom panel). The order parameter $P_n$ is shown for different values of $n=20000,10000,5000,2500$ indicated with solid (red), dashed (green) dot-dashed (orange) and dotted (blue) thick lines. The predicted scaling of the order parameter in the infinite network limit is indicated with the thin dashed (black) line. }
\label{Fig:transition}  
\end{figure}
\subsection{Critical scaling of the order parameter}

In this paragraph we derive the asymptotic behavior of the order parameter $P_{\infty}$ given by Eq.\  (\ref{Pinf2}) close to the percolation threshold $p_c=0$.
By  approximating $M_n$ with Eq.\ (\ref{Mp4}) the order parameter $P_{\infty}$ given by Eq.\  (\ref{Pinf2}) can be easily shown to obey for 
$0<p\ll1$
\bea
P_{\infty}&\propto&p\lim_{n\to \infty}\frac{1}{\overline{N}_n}\prod_{n'=1}^{n}\lambda_{n'}\nonumber \\
&= &p\exp\left[-\ln\avg{m}\int_0^{\infty}dn(1-\psi_n)\right] 
\label{P3}
\eea
where $\psi_n$ is  defined as \bea
\psi_n=\frac{\ln \lambda_n}{\ln \avg{m}}.
\eea

Using for $\lambda_n$ the expansion given by Eq.\ (\ref{l2})  $\psi_n$ can be expanded to give
\bea
\psi_n=\frac{\ln \lambda_n}{\ln \avg{m}} =\frac{\ln 2}{\ln \avg{m}}+\frac{y_n}{\ln \avg{m}}+O(y_n^2).
\eea
Therefore in the continuous limit for $n$ we get 
\bea
1-\psi\simeq 1-\frac{\ln 2}{\ln \avg{m}}-\frac{y}{\ln m} 
\eea
with the function $y(n)$ given by the scaling function Eq.\  (\ref{scaling}) and diverging for  $n={n_c}$. At a value of $n\sim n_c$,  $T_n$ jumps to $T_n=1$, $\lambda_n=\Avg{m}$.
Consequently we have that $1-\psi_n$ will also have a discontinuity  at $n_c$, i.e.,
\bea
1-\psi_n=\left\{\begin{array}{ccc}f_{\psi}(\hat{n}/n_c)&\mbox {for} &n<n_c \\
0&\mbox{for}& n>n_c\end{array}.\right.
\eea
where $f_\psi(x)$ is a scaling function.
Using this expression in Eq.\  (\ref{P3}) we obtain
\bea
P_{\infty}& \propto &p\exp\left[-\ln \avg{m}\int_0^{\infty} dn (1-\psi_n)\right]\nonumber \\
&= & p\exp\left[-\ln \avg{m}\int_0^{\hat{n}_c}d\hat{n} f_{\psi}(\hat{n}/n_c)\right]
\eea
By changing the variable of integration from $n$ to $x=n/n_c$ we obtain
\bea
P_{\infty}&\propto & p\exp\left[-{n_c}\ln \avg{m}\int_0^{1}dx f_{\psi}(x)\right]
\eea
Finally using the expression for $n_c$ given by Eq.\  (\ref{nc}), by indicating with $\alpha$ the constant
\bea
\alpha=\frac{\ln \avg{m}}{(\overline{m}-2)q_{\overline{m}}}\int_{0}^1 f_{\psi}(x)dx,
\eea
we obtain
\bea
P_{\infty}\propto p\exp\left(-\frac{\alpha}{|\ln (1-p)|^{\overline{m}-2}}\right).
\eea
 Because in the critical region  $p\ll1$, it follows that $P_{\infty}$ follows the asymptotic scaling
\bea
P_{\infty}\propto p\exp(-{\alpha}/{p^{\overline{m}-2}}).
\label{asym}
\eea
This scaling can be validated by numerically integrating Eq.\  (\ref{Vprimeps}) and using the finite size scaling of $P_n$ defined as the fraction of nodes in the giant component of a pseudo-fractal cell complexes evolved up to iteration $n$, i.e.,
\bea
P_n=\frac{M_n}{\overline{N}_n}.
\eea
Our numerical results shown in Fig.\  \ref{Fig:transition} clearly demonstrates that if $n>n_c$ (where $n_c$ is a function of $p$ defined by Eq.\  (\ref{nc})) then $P_n$ follows the asymptotic scaling defined in Eq.\  (\ref{asym}). However if $n<n_c$, then $P_n$ saturates to a constant value. This phenomenology is  in perfect agreement with  our theoretical understanding of the critical properties of link percolation on pseudo-fractal cell complexes. 
\section{Conclusions}
In this work we have studied the nature of the link percolation transition in pseudo-fractal simplicial and cell complexes. The pseudo-fractal generalized networks under study include deterministic and random cell complexes, made by gluing together $m$-polygons with the same number of sides $m$ or with random number of sides $m$ drawn from a $q_m$ distribution.
All these generalized network topologies display a link percolation transition at $p_c=0$. However the critical behavior of the order parameter depends on the topology of the generalized network structure.
For deterministic pseudo-fractal simplicial complexes ($m=3$) we confirm the results of Ref.\  \cite{doro3} showing that the order parameter is exponentially suppressed by a term $1/p$ and we predict an additional modulation of the order parameter by a factor $p$.    
For deterministic pseudo-fractal cell complexes with $m>3$ we show that the exponential suppression is more severe than for simplicial complexes and decays as $1/p^{m-2}$.
Finally for random cell complexes we show that the critical behavior is dominated by the smallest value of $m$, $\overline{m}$ for which $q_m>0$.
This work shows clearly that the dynamical processes defined on simplicial complexes and their cell complex counterpart might be significantly different, emphasizing the important role that network topology and geometry have on dynamical processes. 
\bibliographystyle{apsrev4-1}
\bibliography{biblio}

\begin{thebibliography}{42}%
\makeatletter
\providecommand \@ifxundefined [1]{%
 \@ifx{#1\undefined}
}%
\providecommand \@ifnum [1]{%
 \ifnum #1\expandafter \@firstoftwo
 \else \expandafter \@secondoftwo
 \fi
}%
\providecommand \@ifx [1]{%
 \ifx #1\expandafter \@firstoftwo
 \else \expandafter \@secondoftwo
 \fi
}%
\providecommand \natexlab [1]{#1}%
\providecommand \enquote  [1]{``#1''}%
\providecommand \bibnamefont  [1]{#1}%
\providecommand \bibfnamefont [1]{#1}%
\providecommand \citenamefont [1]{#1}%
\providecommand \href@noop [0]{\@secondoftwo}%
\providecommand \href [0]{\begingroup \@sanitize@url \@href}%
\providecommand \@href[1]{\@@startlink{#1}\@@href}%
\providecommand \@@href[1]{\endgroup#1\@@endlink}%
\providecommand \@sanitize@url [0]{\catcode `\\12\catcode `\$12\catcode
  `\&12\catcode `\#12\catcode `\^12\catcode `\_12\catcode `\%12\relax}%
\providecommand \@@startlink[1]{}%
\providecommand \@@endlink[0]{}%
\providecommand \url  [0]{\begingroup\@sanitize@url \@url }%
\providecommand \@url [1]{\endgroup\@href {#1}{\urlprefix }}%
\providecommand \urlprefix  [0]{URL }%
\providecommand \Eprint [0]{\href }%
\providecommand \doibase [0]{http://dx.doi.org/}%
\providecommand \selectlanguage [0]{\@gobble}%
\providecommand \bibinfo  [0]{\@secondoftwo}%
\providecommand \bibfield  [0]{\@secondoftwo}%
\providecommand \translation [1]{[#1]}%
\providecommand \BibitemOpen [0]{}%
\providecommand \bibitemStop [0]{}%
\providecommand \bibitemNoStop [0]{.\EOS\space}%
\providecommand \EOS [0]{\spacefactor3000\relax}%
\providecommand \BibitemShut  [1]{\csname bibitem#1\endcsname}%
\let\auto@bib@innerbib\@empty
\bibitem [{\citenamefont {Bianconi}(2015)}]{Perspectives}%
  \BibitemOpen
  \bibfield  {author} {\bibinfo {author} {\bibfnamefont {G.}~\bibnamefont
  {Bianconi}},\ }\href@noop {} {\bibfield  {journal} {\bibinfo  {journal} {EPL
  (Europhys. Lett.)}\ }\textbf {\bibinfo {volume} {111}},\ \bibinfo {pages}
  {56001} (\bibinfo {year} {2015})}\BibitemShut {NoStop}%
\bibitem [{\citenamefont {Salnikov}\ \emph {et~al.}(2018)\citenamefont
  {Salnikov}, \citenamefont {Cassese},\ and\ \citenamefont
  {Lambiotte}}]{Lambiotte}%
  \BibitemOpen
  \bibfield  {author} {\bibinfo {author} {\bibfnamefont {V.}~\bibnamefont
  {Salnikov}}, \bibinfo {author} {\bibfnamefont {D.}~\bibnamefont {Cassese}}, \
  and\ \bibinfo {author} {\bibfnamefont {R.}~\bibnamefont {Lambiotte}},\
  }\href@noop {} {\bibfield  {journal} {\bibinfo  {journal} {Euro. J. Phys.}\
  }\textbf {\bibinfo {volume} {40}},\ \bibinfo {pages} {014001} (\bibinfo
  {year} {2018})}\BibitemShut {NoStop}%
\bibitem [{\citenamefont {Giusti}\ \emph {et~al.}(2016)\citenamefont {Giusti},
  \citenamefont {Ghrist},\ and\ \citenamefont {Bassett}}]{Bassett}%
  \BibitemOpen
  \bibfield  {author} {\bibinfo {author} {\bibfnamefont {C.}~\bibnamefont
  {Giusti}}, \bibinfo {author} {\bibfnamefont {R.}~\bibnamefont {Ghrist}}, \
  and\ \bibinfo {author} {\bibfnamefont {D.~S.}\ \bibnamefont {Bassett}},\
  }\href@noop {} {\bibfield  {journal} {\bibinfo  {journal} {J. Computational
  Neuroscience}\ }\textbf {\bibinfo {volume} {41}},\ \bibinfo {pages} {1}
  (\bibinfo {year} {2016})}\BibitemShut {NoStop}%
\bibitem [{\citenamefont {Petri}\ \emph {et~al.}(2014)\citenamefont {Petri},
  \citenamefont {Expert}, \citenamefont {Turkheimer}, \citenamefont
  {Carhart-Harris}, \citenamefont {Nutt}, \citenamefont {Hellyer},\ and\
  \citenamefont {Vaccarino}}]{Petri}%
  \BibitemOpen
  \bibfield  {author} {\bibinfo {author} {\bibfnamefont {G.}~\bibnamefont
  {Petri}}, \bibinfo {author} {\bibfnamefont {P.}~\bibnamefont {Expert}},
  \bibinfo {author} {\bibfnamefont {F.}~\bibnamefont {Turkheimer}}, \bibinfo
  {author} {\bibfnamefont {R.}~\bibnamefont {Carhart-Harris}}, \bibinfo
  {author} {\bibfnamefont {D.}~\bibnamefont {Nutt}}, \bibinfo {author}
  {\bibfnamefont {P.~J.}\ \bibnamefont {Hellyer}}, \ and\ \bibinfo {author}
  {\bibfnamefont {F.}~\bibnamefont {Vaccarino}},\ }\href@noop {} {\bibfield
  {journal} {\bibinfo  {journal} {J. Royal Society Interface}\ }\textbf
  {\bibinfo {volume} {11}},\ \bibinfo {pages} {20140873} (\bibinfo {year}
  {2014})}\BibitemShut {NoStop}%
\bibitem [{\citenamefont {Reimann}\ \emph {et~al.}(2017)\citenamefont
  {Reimann}, \citenamefont {Nolte}, \citenamefont {Scolamiero}, \citenamefont
  {Turner}, \citenamefont {Perin}, \citenamefont {Chindemi}, \citenamefont
  {D{\l}otko}, \citenamefont {Levi}, \citenamefont {Hess},\ and\ \citenamefont
  {Markram}}]{BlueBrain}%
  \BibitemOpen
  \bibfield  {author} {\bibinfo {author} {\bibfnamefont {M.~W.}\ \bibnamefont
  {Reimann}}, \bibinfo {author} {\bibfnamefont {M.}~\bibnamefont {Nolte}},
  \bibinfo {author} {\bibfnamefont {M.}~\bibnamefont {Scolamiero}}, \bibinfo
  {author} {\bibfnamefont {K.}~\bibnamefont {Turner}}, \bibinfo {author}
  {\bibfnamefont {R.}~\bibnamefont {Perin}}, \bibinfo {author} {\bibfnamefont
  {G.}~\bibnamefont {Chindemi}}, \bibinfo {author} {\bibfnamefont
  {P.}~\bibnamefont {D{\l}otko}}, \bibinfo {author} {\bibfnamefont
  {R.}~\bibnamefont {Levi}}, \bibinfo {author} {\bibfnamefont {K.}~\bibnamefont
  {Hess}}, \ and\ \bibinfo {author} {\bibfnamefont {H.}~\bibnamefont
  {Markram}},\ }\href@noop {} {\bibfield  {journal} {\bibinfo  {journal}
  {Frontiers in Computational Neuroscience}\ }\textbf {\bibinfo {volume}
  {11}},\ \bibinfo {pages} {48} (\bibinfo {year} {2017})}\BibitemShut {NoStop}%
\bibitem [{\citenamefont {Petri}\ and\ \citenamefont {Barrat}(2018)}]{Alain}%
  \BibitemOpen
  \bibfield  {author} {\bibinfo {author} {\bibfnamefont {G.}~\bibnamefont
  {Petri}}\ and\ \bibinfo {author} {\bibfnamefont {A.}~\bibnamefont {Barrat}},\
  }\href@noop {} {\bibfield  {journal} {\bibinfo  {journal} {Phys. Rev. Lett.}\
  }\textbf {\bibinfo {volume} {121}},\ \bibinfo {pages} {228301} (\bibinfo
  {year} {2018})}\BibitemShut {NoStop}%
\bibitem [{\citenamefont {Iacopini}\ \emph {et~al.}(2019)\citenamefont
  {Iacopini}, \citenamefont {Petri}, \citenamefont {Barrat},\ and\
  \citenamefont {Latora}}]{iacopini2019simplicial}%
  \BibitemOpen
  \bibfield  {author} {\bibinfo {author} {\bibfnamefont {I.}~\bibnamefont
  {Iacopini}}, \bibinfo {author} {\bibfnamefont {G.}~\bibnamefont {Petri}},
  \bibinfo {author} {\bibfnamefont {A.}~\bibnamefont {Barrat}}, \ and\ \bibinfo
  {author} {\bibfnamefont {V.}~\bibnamefont {Latora}},\ }\href@noop {}
  {\bibfield  {journal} {\bibinfo  {journal} {Nat. Comm.}\ }\textbf {\bibinfo
  {volume} {10}},\ \bibinfo {pages} {2485} (\bibinfo {year}
  {2019})}\BibitemShut {NoStop}%
\bibitem [{\citenamefont {Alvarez-Rodriguez}\ \emph {et~al.}(2020)\citenamefont
  {Alvarez-Rodriguez}, \citenamefont {Battiston}, \citenamefont {de~Arruda},
  \citenamefont {Moreno}, \citenamefont {Perc},\ and\ \citenamefont
  {Latora}}]{Yamir}%
  \BibitemOpen
  \bibfield  {author} {\bibinfo {author} {\bibfnamefont {U.}~\bibnamefont
  {Alvarez-Rodriguez}}, \bibinfo {author} {\bibfnamefont {F.}~\bibnamefont
  {Battiston}}, \bibinfo {author} {\bibfnamefont {G.~F.}\ \bibnamefont
  {de~Arruda}}, \bibinfo {author} {\bibfnamefont {Y.}~\bibnamefont {Moreno}},
  \bibinfo {author} {\bibfnamefont {M.}~\bibnamefont {Perc}}, \ and\ \bibinfo
  {author} {\bibfnamefont {V.}~\bibnamefont {Latora}},\ }\href@noop {}
  {\bibfield  {journal} {\bibinfo  {journal} {arXiv preprint arXiv:2001.10313}\
  } (\bibinfo {year} {2020})}\BibitemShut {NoStop}%
\bibitem [{\citenamefont {Papadopoulos}\ \emph {et~al.}(2018)\citenamefont
  {Papadopoulos}, \citenamefont {Porter}, \citenamefont {Daniels},\ and\
  \citenamefont {Bassett}}]{Bassett_granular}%
  \BibitemOpen
  \bibfield  {author} {\bibinfo {author} {\bibfnamefont {L.}~\bibnamefont
  {Papadopoulos}}, \bibinfo {author} {\bibfnamefont {M.~A.}\ \bibnamefont
  {Porter}}, \bibinfo {author} {\bibfnamefont {K.~E.}\ \bibnamefont {Daniels}},
  \ and\ \bibinfo {author} {\bibfnamefont {D.~S.}\ \bibnamefont {Bassett}},\
  }\href@noop {} {\bibfield  {journal} {\bibinfo  {journal} {Journal of Complex
  Networks}\ }\textbf {\bibinfo {volume} {6}},\ \bibinfo {pages} {485}
  (\bibinfo {year} {2018})}\BibitemShut {NoStop}%
\bibitem [{\citenamefont {{\v{S}}uvakov}\ \emph {et~al.}(2018)\citenamefont
  {{\v{S}}uvakov}, \citenamefont {Andjelkovi{\'c}},\ and\ \citenamefont
  {Tadi{\'c}}}]{Tadic}%
  \BibitemOpen
  \bibfield  {author} {\bibinfo {author} {\bibfnamefont {M.}~\bibnamefont
  {{\v{S}}uvakov}}, \bibinfo {author} {\bibfnamefont {M.}~\bibnamefont
  {Andjelkovi{\'c}}}, \ and\ \bibinfo {author} {\bibfnamefont {B.}~\bibnamefont
  {Tadi{\'c}}},\ }\href@noop {} {\bibfield  {journal} {\bibinfo  {journal}
  {Scientific Reports}\ }\textbf {\bibinfo {volume} {8}},\ \bibinfo {pages} {1}
  (\bibinfo {year} {2018})}\BibitemShut {NoStop}%
\bibitem [{\citenamefont {Bianconi}\ and\ \citenamefont
  {Rahmede}(2017{\natexlab{a}})}]{Emergent}%
  \BibitemOpen
  \bibfield  {author} {\bibinfo {author} {\bibfnamefont {G.}~\bibnamefont
  {Bianconi}}\ and\ \bibinfo {author} {\bibfnamefont {C.}~\bibnamefont
  {Rahmede}},\ }\href@noop {} {\bibfield  {journal} {\bibinfo  {journal}
  {Scientific Reports}\ }\textbf {\bibinfo {volume} {7}},\ \bibinfo {pages}
  {41974} (\bibinfo {year} {2017}{\natexlab{a}})}\BibitemShut {NoStop}%
\bibitem [{\citenamefont {Bianconi}\ and\ \citenamefont
  {Rahmede}(2017{\natexlab{b}})}]{Hyperbolic}%
  \BibitemOpen
  \bibfield  {author} {\bibinfo {author} {\bibfnamefont {G.}~\bibnamefont
  {Bianconi}}\ and\ \bibinfo {author} {\bibfnamefont {C.}~\bibnamefont
  {Rahmede}},\ }\href@noop {} {\bibfield  {journal} {\bibinfo  {journal}
  {Scientific Reports}\ }\textbf {\bibinfo {volume} {7}},\ \bibinfo {pages}
  {41974} (\bibinfo {year} {2017}{\natexlab{b}})}\BibitemShut {NoStop}%
\bibitem [{\citenamefont {Mulder}\ and\ \citenamefont
  {Bianconi}(2018)}]{Polytopes}%
  \BibitemOpen
  \bibfield  {author} {\bibinfo {author} {\bibfnamefont {D.}~\bibnamefont
  {Mulder}}\ and\ \bibinfo {author} {\bibfnamefont {G.}~\bibnamefont
  {Bianconi}},\ }\href@noop {} {\bibfield  {journal} {\bibinfo  {journal} {J.
  Stat. Phys.}\ }\textbf {\bibinfo {volume} {173}},\ \bibinfo {pages} {783}
  (\bibinfo {year} {2018})}\BibitemShut {NoStop}%
\bibitem [{\citenamefont {Bianconi}\ and\ \citenamefont
  {Ziff}(2018)}]{Bianconi_Ziff}%
  \BibitemOpen
  \bibfield  {author} {\bibinfo {author} {\bibfnamefont {G.}~\bibnamefont
  {Bianconi}}\ and\ \bibinfo {author} {\bibfnamefont {R.~M.}\ \bibnamefont
  {Ziff}},\ }\href@noop {} {\bibfield  {journal} {\bibinfo  {journal} {Phys.
  Rev. E}\ }\textbf {\bibinfo {volume} {98}},\ \bibinfo {pages} {052308}
  (\bibinfo {year} {2018})}\BibitemShut {NoStop}%
\bibitem [{\citenamefont {Kryven}\ \emph {et~al.}(2019)\citenamefont {Kryven},
  \citenamefont {Ziff},\ and\ \citenamefont
  {Bianconi}}]{kryven2019renormalization}%
  \BibitemOpen
  \bibfield  {author} {\bibinfo {author} {\bibfnamefont {I.}~\bibnamefont
  {Kryven}}, \bibinfo {author} {\bibfnamefont {R.~M.}\ \bibnamefont {Ziff}}, \
  and\ \bibinfo {author} {\bibfnamefont {G.}~\bibnamefont {Bianconi}},\
  }\href@noop {} {\bibfield  {journal} {\bibinfo  {journal} {Physical Review
  E}\ }\textbf {\bibinfo {volume} {100}},\ \bibinfo {pages} {022306} (\bibinfo
  {year} {2019})}\BibitemShut {NoStop}%
\bibitem [{\citenamefont {Bianconi}\ \emph {et~al.}(2019)\citenamefont
  {Bianconi}, \citenamefont {Kryven},\ and\ \citenamefont {Ziff}}]{branching}%
  \BibitemOpen
  \bibfield  {author} {\bibinfo {author} {\bibfnamefont {G.}~\bibnamefont
  {Bianconi}}, \bibinfo {author} {\bibfnamefont {I.}~\bibnamefont {Kryven}}, \
  and\ \bibinfo {author} {\bibfnamefont {R.~M.}\ \bibnamefont {Ziff}},\
  }\href@noop {} {\bibfield  {journal} {\bibinfo  {journal} {Phys. Rev. E}\
  }\textbf {\bibinfo {volume} {100}},\ \bibinfo {pages} {062311} (\bibinfo
  {year} {2019})}\BibitemShut {NoStop}%
\bibitem [{\citenamefont {Hasegawa}\ and\ \citenamefont
  {Nemoto}(2013)}]{hasegawa}%
  \BibitemOpen
  \bibfield  {author} {\bibinfo {author} {\bibfnamefont {T.}~\bibnamefont
  {Hasegawa}}\ and\ \bibinfo {author} {\bibfnamefont {K.}~\bibnamefont
  {Nemoto}},\ }\href@noop {} {\bibfield  {journal} {\bibinfo  {journal} {Phys.
  Rev. E}\ }\textbf {\bibinfo {volume} {88}},\ \bibinfo {pages} {062807}
  (\bibinfo {year} {2013})}\BibitemShut {NoStop}%
\bibitem [{\citenamefont {Skardal}\ and\ \citenamefont
  {Arenas}(2019)}]{Arenas}%
  \BibitemOpen
  \bibfield  {author} {\bibinfo {author} {\bibfnamefont {P.~S.}\ \bibnamefont
  {Skardal}}\ and\ \bibinfo {author} {\bibfnamefont {A.}~\bibnamefont
  {Arenas}},\ }\href@noop {} {\bibfield  {journal} {\bibinfo  {journal} {arXiv
  preprint arXiv:1903.12131}\ } (\bibinfo {year} {2019})}\BibitemShut {NoStop}%
\bibitem [{\citenamefont {Mill{\'a}n}\ \emph {et~al.}(2020)\citenamefont
  {Mill{\'a}n}, \citenamefont {Torres},\ and\ \citenamefont
  {Bianconi}}]{millan2020explosive}%
  \BibitemOpen
  \bibfield  {author} {\bibinfo {author} {\bibfnamefont {A.~P.}\ \bibnamefont
  {Mill{\'a}n}}, \bibinfo {author} {\bibfnamefont {J.~J.}\ \bibnamefont
  {Torres}}, \ and\ \bibinfo {author} {\bibfnamefont {G.}~\bibnamefont
  {Bianconi}},\ }\href@noop {} {\bibfield  {journal} {\bibinfo  {journal}
  {Phys. Rev. Lett.}\ }\textbf {\bibinfo {volume} {124}},\ \bibinfo {pages}
  {218301} (\bibinfo {year} {2020})}\BibitemShut {NoStop}%
\bibitem [{\citenamefont {Gambuzza}\ \emph {et~al.}(2020)\citenamefont
  {Gambuzza}, \citenamefont {Di~Patti}, \citenamefont {Gallo}, \citenamefont
  {Lepri}, \citenamefont {Romance}, \citenamefont {Criado}, \citenamefont
  {Frasca}, \citenamefont {Latora},\ and\ \citenamefont
  {Boccaletti}}]{Boccaletti}%
  \BibitemOpen
  \bibfield  {author} {\bibinfo {author} {\bibfnamefont {L.}~\bibnamefont
  {Gambuzza}}, \bibinfo {author} {\bibfnamefont {F.}~\bibnamefont {Di~Patti}},
  \bibinfo {author} {\bibfnamefont {L.}~\bibnamefont {Gallo}}, \bibinfo
  {author} {\bibfnamefont {S.}~\bibnamefont {Lepri}}, \bibinfo {author}
  {\bibfnamefont {M.}~\bibnamefont {Romance}}, \bibinfo {author} {\bibfnamefont
  {R.}~\bibnamefont {Criado}}, \bibinfo {author} {\bibfnamefont
  {M.}~\bibnamefont {Frasca}}, \bibinfo {author} {\bibfnamefont
  {V.}~\bibnamefont {Latora}}, \ and\ \bibinfo {author} {\bibfnamefont
  {S.}~\bibnamefont {Boccaletti}},\ }\href@noop {} {\bibfield  {journal}
  {\bibinfo  {journal} {arXiv preprint arXiv:2004.03913}\ } (\bibinfo {year}
  {2020})}\BibitemShut {NoStop}%
\bibitem [{\citenamefont {Lucas}\ \emph {et~al.}(2020)\citenamefont {Lucas},
  \citenamefont {Cencetti},\ and\ \citenamefont
  {Battiston}}]{Battiston_synchronization}%
  \BibitemOpen
  \bibfield  {author} {\bibinfo {author} {\bibfnamefont {M.}~\bibnamefont
  {Lucas}}, \bibinfo {author} {\bibfnamefont {G.}~\bibnamefont {Cencetti}}, \
  and\ \bibinfo {author} {\bibfnamefont {F.}~\bibnamefont {Battiston}},\
  }\href@noop {} {\bibfield  {journal} {\bibinfo  {journal} {arXiv preprint
  arXiv:2003.09734}\ } (\bibinfo {year} {2020})}\BibitemShut {NoStop}%
\bibitem [{\citenamefont {Mill{\'a}n}\ \emph {et~al.}(2018)\citenamefont
  {Mill{\'a}n}, \citenamefont {Torres},\ and\ \citenamefont
  {Bianconi}}]{millan2018complex}%
  \BibitemOpen
  \bibfield  {author} {\bibinfo {author} {\bibfnamefont {A.~P.}\ \bibnamefont
  {Mill{\'a}n}}, \bibinfo {author} {\bibfnamefont {J.~J.}\ \bibnamefont
  {Torres}}, \ and\ \bibinfo {author} {\bibfnamefont {G.}~\bibnamefont
  {Bianconi}},\ }\href@noop {} {\bibfield  {journal} {\bibinfo  {journal}
  {Scientific Reports}\ }\textbf {\bibinfo {volume} {8}},\ \bibinfo {pages}
  {9910} (\bibinfo {year} {2018})}\BibitemShut {NoStop}%
\bibitem [{\citenamefont {Mill{\'a}n}\ \emph {et~al.}(2019)\citenamefont
  {Mill{\'a}n}, \citenamefont {Torres},\ and\ \citenamefont
  {Bianconi}}]{millan2019synchronization}%
  \BibitemOpen
  \bibfield  {author} {\bibinfo {author} {\bibfnamefont {A.~P.}\ \bibnamefont
  {Mill{\'a}n}}, \bibinfo {author} {\bibfnamefont {J.~J.}\ \bibnamefont
  {Torres}}, \ and\ \bibinfo {author} {\bibfnamefont {G.}~\bibnamefont
  {Bianconi}},\ }\href@noop {} {\bibfield  {journal} {\bibinfo  {journal}
  {Phys. Rev. E}\ }\textbf {\bibinfo {volume} {99}},\ \bibinfo {pages} {022307}
  (\bibinfo {year} {2019})}\BibitemShut {NoStop}%
\bibitem [{\citenamefont {St-Onge}\ \emph {et~al.}(2020)\citenamefont
  {St-Onge}, \citenamefont {Thibeault}, \citenamefont {Allard}, \citenamefont
  {Dub{\ 'e}},\ and\ \citenamefont {H{\'e}bert-Dufresne}}]{laurent}%
  \BibitemOpen
  \bibfield  {author} {\bibinfo {author} {\bibfnamefont {G.}~\bibnamefont
  {St-Onge}}, \bibinfo {author} {\bibfnamefont {V.}~\bibnamefont {Thibeault}},
  \bibinfo {author} {\bibfnamefont {A.}~\bibnamefont {Allard}}, \bibinfo
  {author} {\bibfnamefont {L.~J.}\ \bibnamefont {Dub{\ 'e}}}, \ and\ \bibinfo
  {author} {\bibnamefont {H{\'e}bert-Dufresne}},\ }\href@noop {} {\bibfield
  {journal} {\bibinfo  {journal} {arXiv preprint arXiv:2004.10203}\ } (\bibinfo
  {year} {2020})}\BibitemShut {NoStop}%
\bibitem [{\citenamefont {Bianconi}\ and\ \citenamefont
  {Dorogovstev}(2020)}]{bianconi2020spectral}%
  \BibitemOpen
  \bibfield  {author} {\bibinfo {author} {\bibfnamefont {G.}~\bibnamefont
  {Bianconi}}\ and\ \bibinfo {author} {\bibfnamefont {S.~N.}\ \bibnamefont
  {Dorogovstev}},\ }\href@noop {} {\bibfield  {journal} {\bibinfo  {journal}
  {J. Stat. Mech.: Th. Exp.}\ }\textbf {\bibinfo {volume} {2020}},\ \bibinfo
  {pages} {014005} (\bibinfo {year} {2020})}\BibitemShut {NoStop}%
\bibitem [{\citenamefont {Reitz}\ and\ \citenamefont {Bianconi}(2020)}]{reitz}%
  \BibitemOpen
  \bibfield  {author} {\bibinfo {author} {\bibfnamefont {M.}~\bibnamefont
  {Reitz}}\ and\ \bibinfo {author} {\bibfnamefont {G.}~\bibnamefont
  {Bianconi}},\ }\href@noop {} {\bibfield  {journal} {\bibinfo  {journal}
  {arXiv preprint arXiv:2003.09143}\ } (\bibinfo {year} {2020})}\BibitemShut
  {NoStop}%
\bibitem [{\citenamefont {Torres}\ and\ \citenamefont
  {Bianconi}(2020)}]{torres}%
  \BibitemOpen
  \bibfield  {author} {\bibinfo {author} {\bibfnamefont {J.~J.}\ \bibnamefont
  {Torres}}\ and\ \bibinfo {author} {\bibfnamefont {G.}~\bibnamefont
  {Bianconi}},\ }\href@noop {} {\bibfield  {journal} {\bibinfo  {journal}
  {JPhys. Complexity}\ }\textbf {\bibinfo {volume} {1}},\ \bibinfo {pages}
  {015002} (\bibinfo {year} {2020})}\BibitemShut {NoStop}%
\bibitem [{\citenamefont {Carletti}\ \emph {et~al.}(2020)\citenamefont
  {Carletti}, \citenamefont {Battiston}, \citenamefont {Cencetti},\ and\
  \citenamefont {Fanelli}}]{Battiston_RW}%
  \BibitemOpen
  \bibfield  {author} {\bibinfo {author} {\bibfnamefont {T.}~\bibnamefont
  {Carletti}}, \bibinfo {author} {\bibfnamefont {F.}~\bibnamefont {Battiston}},
  \bibinfo {author} {\bibfnamefont {G.}~\bibnamefont {Cencetti}}, \ and\
  \bibinfo {author} {\bibfnamefont {D.}~\bibnamefont {Fanelli}},\ }\href@noop
  {} {\bibfield  {journal} {\bibinfo  {journal} {Phys. Rev. E}\ }\textbf
  {\bibinfo {volume} {101}},\ \bibinfo {pages} {022308} (\bibinfo {year}
  {2020})}\BibitemShut {NoStop}%
\bibitem [{\citenamefont {Migdal}(1976)}]{migdal1976phase}%
  \BibitemOpen
  \bibfield  {author} {\bibinfo {author} {\bibfnamefont {A.~A.}\ \bibnamefont
  {Migdal}},\ }\href@noop {} {\bibfield  {journal} {\bibinfo  {journal} {Soviet
  Journal of Experimental and Theoretical Physics}\ }\textbf {\bibinfo {volume}
  {42}},\ \bibinfo {pages} {743} (\bibinfo {year} {1976})}\BibitemShut
  {NoStop}%
\bibitem [{\citenamefont {Kadanoff}(1976)}]{kadanoff1976notes}%
  \BibitemOpen
  \bibfield  {author} {\bibinfo {author} {\bibfnamefont {L.~P.}\ \bibnamefont
  {Kadanoff}},\ }\href@noop {} {\bibfield  {journal} {\bibinfo  {journal}
  {Annals of Physics}\ }\textbf {\bibinfo {volume} {100}},\ \bibinfo {pages}
  {359} (\bibinfo {year} {1976})}\BibitemShut {NoStop}%
\bibitem [{\citenamefont {Boettcher}\ \emph {et~al.}(2012)\citenamefont
  {Boettcher}, \citenamefont {Singh},\ and\ \citenamefont
  {Ziff}}]{hyperbolic_Ziff}%
  \BibitemOpen
  \bibfield  {author} {\bibinfo {author} {\bibfnamefont {S.}~\bibnamefont
  {Boettcher}}, \bibinfo {author} {\bibfnamefont {V.}~\bibnamefont {Singh}}, \
  and\ \bibinfo {author} {\bibfnamefont {R.~M.}\ \bibnamefont {Ziff}},\
  }\href@noop {} {\bibfield  {journal} {\bibinfo  {journal} {Nat. Comm.}\
  }\textbf {\bibinfo {volume} {3}},\ \bibinfo {pages} {787} (\bibinfo {year}
  {2012})}\BibitemShut {NoStop}%
\bibitem [{\citenamefont {Boettcher}\ \emph {et~al.}(2009)\citenamefont
  {Boettcher}, \citenamefont {Cook},\ and\ \citenamefont {Ziff}}]{Patchy}%
  \BibitemOpen
  \bibfield  {author} {\bibinfo {author} {\bibfnamefont {S.}~\bibnamefont
  {Boettcher}}, \bibinfo {author} {\bibfnamefont {J.~L.}\ \bibnamefont {Cook}},
  \ and\ \bibinfo {author} {\bibfnamefont {R.~M.}\ \bibnamefont {Ziff}},\
  }\href@noop {} {\bibfield  {journal} {\bibinfo  {journal} {Phys. Rev. E}\
  }\textbf {\bibinfo {volume} {80}},\ \bibinfo {pages} {041115} (\bibinfo
  {year} {2009})}\BibitemShut {NoStop}%
\bibitem [{\citenamefont {Nogawa}(2018)}]{RG}%
  \BibitemOpen
  \bibfield  {author} {\bibinfo {author} {\bibfnamefont {T.}~\bibnamefont
  {Nogawa}},\ }\href@noop {} {\bibfield  {journal} {\bibinfo  {journal} {J.
  Phys. A: Math. Gen.}\ }\textbf {\bibinfo {volume} {51}},\ \bibinfo {pages}
  {505003} (\bibinfo {year} {2018})}\BibitemShut {NoStop}%
\bibitem [{\citenamefont {Hasegawa}\ \emph {et~al.}(2010)\citenamefont
  {Hasegawa}, \citenamefont {Sato},\ and\ \citenamefont {Nemoto}}]{flower_tau}%
  \BibitemOpen
  \bibfield  {author} {\bibinfo {author} {\bibfnamefont {T.}~\bibnamefont
  {Hasegawa}}, \bibinfo {author} {\bibfnamefont {M.}~\bibnamefont {Sato}}, \
  and\ \bibinfo {author} {\bibfnamefont {K.}~\bibnamefont {Nemoto}},\
  }\href@noop {} {\bibfield  {journal} {\bibinfo  {journal} {Phys. Rev. E}\
  }\textbf {\bibinfo {volume} {82}},\ \bibinfo {pages} {046101} (\bibinfo
  {year} {2010})}\BibitemShut {NoStop}%
\bibitem [{\citenamefont {Nogawa}\ and\ \citenamefont
  {Hasegawa}(2014)}]{tricritical}%
  \BibitemOpen
  \bibfield  {author} {\bibinfo {author} {\bibfnamefont {T.}~\bibnamefont
  {Nogawa}}\ and\ \bibinfo {author} {\bibfnamefont {T.}~\bibnamefont
  {Hasegawa}},\ }\href@noop {} {\bibfield  {journal} {\bibinfo  {journal}
  {Phys. Rev. E}\ }\textbf {\bibinfo {volume} {89}},\ \bibinfo {pages} {042803}
  (\bibinfo {year} {2014})}\BibitemShut {NoStop}%
\bibitem [{\citenamefont {Auto}\ \emph {et~al.}(2008)\citenamefont {Auto},
  \citenamefont {Moreira}, \citenamefont {Herrmann},\ and\ \citenamefont
  {Andrade~Jr.}}]{percolation_Apollonian}%
  \BibitemOpen
  \bibfield  {author} {\bibinfo {author} {\bibfnamefont {D.~M.}\ \bibnamefont
  {Auto}}, \bibinfo {author} {\bibfnamefont {A.~A.}\ \bibnamefont {Moreira}},
  \bibinfo {author} {\bibfnamefont {H.~J.}\ \bibnamefont {Herrmann}}, \ and\
  \bibinfo {author} {\bibfnamefont {J.~S.}\ \bibnamefont {Andrade~Jr.}},\
  }\href@noop {} {\bibfield  {journal} {\bibinfo  {journal} {Phys. Rev. E}\
  }\textbf {\bibinfo {volume} {78}},\ \bibinfo {pages} {066112} (\bibinfo
  {year} {2008})}\BibitemShut {NoStop}%
\bibitem [{\citenamefont {Boettcher}\ and\ \citenamefont
  {Brunson}(2015)}]{Boettcher_RG}%
  \BibitemOpen
  \bibfield  {author} {\bibinfo {author} {\bibfnamefont {S.}~\bibnamefont
  {Boettcher}}\ and\ \bibinfo {author} {\bibfnamefont {C.~T.}\ \bibnamefont
  {Brunson}},\ }\href@noop {} {\bibfield  {journal} {\bibinfo  {journal} {EPL
  (Europhysics Lett.)}\ }\textbf {\bibinfo {volume} {110}},\ \bibinfo {pages}
  {26005} (\bibinfo {year} {2015})}\BibitemShut {NoStop}%
\bibitem [{\citenamefont {Singh}\ \emph {et~al.}(2014)\citenamefont {Singh},
  \citenamefont {Brunson},\ and\ \citenamefont {Boettcher}}]{Boettcher_Potts}%
  \BibitemOpen
  \bibfield  {author} {\bibinfo {author} {\bibfnamefont {V.}~\bibnamefont
  {Singh}}, \bibinfo {author} {\bibfnamefont {C.~T.}\ \bibnamefont {Brunson}},
  \ and\ \bibinfo {author} {\bibfnamefont {S.}~\bibnamefont {Boettcher}},\
  }\href@noop {} {\bibfield  {journal} {\bibinfo  {journal} {Phys. Rev. E}\
  }\textbf {\bibinfo {volume} {90}},\ \bibinfo {pages} {052119} (\bibinfo
  {year} {2014})}\BibitemShut {NoStop}%
\bibitem [{\citenamefont {Boettcher}\ and\ \citenamefont
  {Brunson}(2011)}]{Boettcher_Ising}%
  \BibitemOpen
  \bibfield  {author} {\bibinfo {author} {\bibfnamefont {S.}~\bibnamefont
  {Boettcher}}\ and\ \bibinfo {author} {\bibfnamefont {C.~T.}\ \bibnamefont
  {Brunson}},\ }\href@noop {} {\bibfield  {journal} {\bibinfo  {journal} {Phys.
  Rev. E}\ }\textbf {\bibinfo {volume} {83}},\ \bibinfo {pages} {021103}
  (\bibinfo {year} {2011})}\BibitemShut {NoStop}%
\bibitem [{\citenamefont {Hinczewski}\ and\ \citenamefont
  {Berker}(2006)}]{Berker_RG}%
  \BibitemOpen
  \bibfield  {author} {\bibinfo {author} {\bibfnamefont {M.}~\bibnamefont
  {Hinczewski}}\ and\ \bibinfo {author} {\bibfnamefont {A.~N.}\ \bibnamefont
  {Berker}},\ }\href@noop {} {\bibfield  {journal} {\bibinfo  {journal} {Phys.
  Rev. E}\ }\textbf {\bibinfo {volume} {73}},\ \bibinfo {pages} {066126}
  (\bibinfo {year} {2006})}\BibitemShut {NoStop}%
\bibitem [{\citenamefont {Dorogovtsev}\ \emph {et~al.}(2002)\citenamefont
  {Dorogovtsev}, \citenamefont {Goltsev},\ and\ \citenamefont
  {Mendes}}]{dorogovtsev2002pseudofractal}%
  \BibitemOpen
  \bibfield  {author} {\bibinfo {author} {\bibfnamefont {S.~N.}\ \bibnamefont
  {Dorogovtsev}}, \bibinfo {author} {\bibfnamefont {A.~V.}\ \bibnamefont
  {Goltsev}}, \ and\ \bibinfo {author} {\bibfnamefont {J.~F.~F.}\ \bibnamefont
  {Mendes}},\ }\href@noop {} {\bibfield  {journal} {\bibinfo  {journal} {Phys.
  Rev. E}\ }\textbf {\bibinfo {volume} {65}},\ \bibinfo {pages} {066122}
  (\bibinfo {year} {2002})}\BibitemShut {NoStop}%
\bibitem [{\citenamefont {Dorogovtsev}(2003)}]{doro3}%
  \BibitemOpen
  \bibfield  {author} {\bibinfo {author} {\bibfnamefont {S.~N.}\ \bibnamefont
  {Dorogovtsev}},\ }\href@noop {} {\bibfield  {journal} {\bibinfo  {journal}
  {Phys. Rev. E}\ }\textbf {\bibinfo {volume} {67}},\ \bibinfo {pages} {045102}
  (\bibinfo {year} {2003})}\BibitemShut {NoStop}%
\end{thebibliography}%
\end{document}